\documentstyle[12pt,epsf]{article}
\newlength{\dinwidth}                                                          
\newlength{\dinmargin}                                                         
\def\3{\ss}
\begin{document}
\vspace{2 cm}
\newcommand{\Gev}       {\mbox{${\rm GeV}$}}
\newcommand{\Gevsq}     {\mbox{${\rm GeV}^2$}}
\newcommand{\qsd}       {\mbox{${Q^2}$}} 
\newcommand{\x}         {\mbox{${\it x}$}}
\newcommand{\smallqsd}  {\mbox{${q^2}$}} 
\newcommand{\ra}        {\mbox{$ \rightarrow $}}
\newcommand {\pom}  {I\hspace{-0.2em}P}
\newcommand {\alphapom} {\mbox{$\alpha_{_{\pom}}$}}
\newcommand {\xpom} {\mbox{$x_{_{\pom}}$}}
\newcommand {\xpomp}[1] {\mbox{$x^{#1}_{_{\pom}}\;$}}
\newcommand {\xpoma} {\mbox{$(1/x_{_{\pom}})^a\;$}}
\def\ctr#1{{\it #1}\\\vspace{10pt}}
\def\si{{\rm si}}
\def\Si{{\rm Si}}
\def\Ci{{\rm Ci}}
\def\qsq{Q^{2}}
\def\yjb{y_{_{JB}}}
\def\xjb{x_{_{JB}}}
\def\qjb{\qsq_{_{JB}}}
\def\gap{\hspace{0.5cm}}
\renewcommand{\thefootnote}{\arabic{footnote}}
\title {
{\bf Measurement of the Diffractive Cross Section 
 in Deep Inelastic Scattering}
\\ 
\author{\rm ZEUS Collaboration \\}}
\date{ }
\maketitle
\vspace{5 cm}
\begin{abstract}

Diffractive scattering of $\gamma^* p \to X + N$, where $N$ is either a
proton or a nucleonic system with $M_N~<~4$~GeV has been measured in deep
inelastic scattering (DIS) at HERA. The cross section was determined by a
novel method as a function of the $\gamma^* p$ c.m. energy $W$ between 60
and 245~GeV and of the mass $M_X$ of the system $X$ up to 15~GeV at
average $Q^2$ values of 14 and 31~GeV$^2$. The diffractive cross section
$d\sigma^{diff} /dM_X$ is, within errors, found to rise linearly with $W$.
Parameterizing the $W$ dependence by the form $d\sigma^{diff}/dM_X \propto
(W^2)^{(2\overline{\alphapom} -2)}$ the DIS data yield for the pomeron
trajectory $\overline{\alphapom} = 1.23 \pm 0.02(stat) \pm 0.04 (syst)$
averaged over $t$ in the measured kinematic range assuming the
longitudinal photon contribution to be zero. This value for the pomeron
trajectory is substantially larger than $\overline{\alphapom}$ extracted
from soft interactions. The value of $\overline{\alphapom}$ measured in
this analysis suggests that a substantial part of the diffractive DIS
cross section originates from processes which can be described by
perturbative QCD. From the measured diffractive cross sections the
diffractive structure function of the proton $F^{D(3)}_2(\beta,Q^2,\xpom)$
has been determined, where $\beta$ is the momentum fraction of the struck
quark in the pomeron. The form $F^{D(3)}_2 = constant \cdot (1/\xpom)^a$
gives a good fit to the data in all $\beta$ and $Q^2$ intervals with $a =
1.46 \pm 0.04 (stat) \pm 0.08(syst)$. 

\end{abstract}

\vspace{-21.5cm}
\begin{flushleft}
\tt DESY 96-018 \\
February 1996 \\
\end{flushleft}

\setcounter{page}{0}
\thispagestyle{empty}   
\newpage
\def\3{\ss}
\footnotesize
\renewcommand{\thepage}{\Roman{page}}
\begin{center}
\begin{large}
The ZEUS Collaboration
\end{large}
\end{center}
M.~Derrick, D.~Krakauer, S.~Magill, D.~Mikunas, B.~Musgrave,
J.R.~Okrasinski, J.~Repond, R.~Stanek, R.L.~Talaga, H.~Zhang \\
{\it Argonne National Laboratory, Argonne, IL, USA}~$^{p}$
% ...
\\[6pt]
\clearpage
\hspace*{1mm}
% ...
$^{ 2}$ now at Univ. of Salerno and INFN Napoli, Italy  \\
\hspace*{1mm}
% ...

\begin{tabular}{lp{15cm}}
$^{a}$ & supported by the Natural Sciences and Engineering Research
         Council of Canada (NSERC) \\
$^{b}$ & supported by the FCAR of Qu\'ebec, Canada\\
$^{c}$ & supported by the German Federal Ministry for Education and
         Science, Research and Technology (BMBF), under contract
         numbers 056BN19I, 056FR19P, 056HH19I, 056HH29I, 056SI79I\\
$^{d}$ & supported by the MINERVA Gesellschaft f\"ur Forschung GmbH,
         and by the Israel Academy of Science \\
$^{e}$ & supported by the German Israeli Foundation, and
         by the Israel Academy of Science \\
$^{f}$ & supported by the Italian National Institute for Nuclear Physics
         (INFN) \\
$^{g}$ & supported by the Japanese Ministry of Education, Science and
         Culture (the Monbusho)
         and its grants for Scientific Research\\
$^{h}$ & supported by the Korean Ministry of Education and Korea Science
         and Engineering Foundation \\
$^{i}$ & supported by the Netherlands Foundation for Research on Matter
         (FOM)\\
$^{j}$ & supported by the Polish State Committee for Scientific
         Research, grants No.~115/E-343/SPUB/P03/109/95, 2P03B 244
         08p02, p03, p04 and p05, and the Foundation for Polish-German
         Collaboration (proj. No. 506/92) \\
$^{k}$ & supported by the Polish State Committee for Scientific
         Research (grant No. 2 P03B 083 08) \\
$^{l}$ & partially supported by the German Federal Ministry for
         Education and Science, Research and Technology (BMBF) \\
$^{m}$ & supported by the German Federal Ministry for Education and
         Science, Research and Technology (BMBF), and the Fund of
         Fundamental Research of Russian Ministry of Science and
         Education and by INTAS-Grant No. 93-63 \\
$^{n}$ & supported by the Spanish Ministry of Education and Science
         through funds provided by CICYT \\
$^{o}$ & supported by the Particle Physics and Astronomy Research
         Council \\
$^{p}$ & supported by the US Department of Energy \\
$^{q}$ & supported by the US National Science Foundation
\end{tabular}

\newpage
\pagenumbering{arabic}
\setcounter{page}{1}
\normalsize

\section{\bf Introduction}
\label{s:intro}

\begin{figure}[ht]
\includegraphics{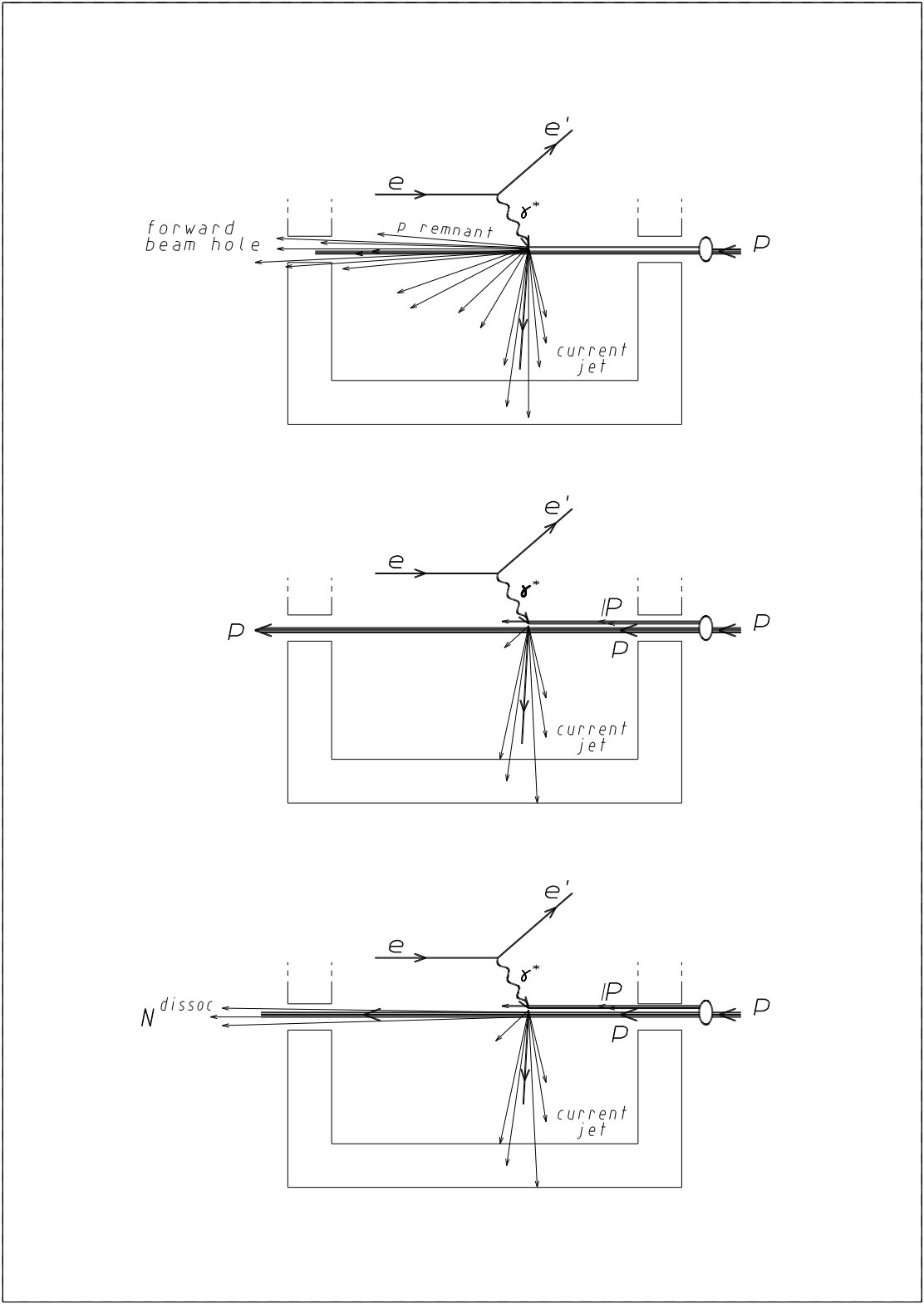}
\unitlength1cm
\begin{picture}(15,20)
\thicklines
\end{picture}
\caption{\label{f:disdiaga}
        { Illustration of deep inelastic scattering:
          nondiffractive scattering (top);
          diffractive scattering without breakup of the proton (middle);
          diffractive scattering with proton dissociation (bottom). 
          }}
\end{figure}

In deep-inelastic electron-proton scattering (DIS), $e^-p \rightarrow e^- $
+ anything (Fig.~\ref{f:disdiaga}), a new class of events was observed by
ZEUS~\cite{Zeplrg,Zeplrgj,Zeplrgf} and H1~\cite{Heplrg} characterized by a
large rapidity gap (LRG) between the direction of the proton beam and the
angle of the first significant energy deposition in the detector.  The
properties of these events indicate a diffractive and leading twist
production mechanism.  The observation of jet production demonstrated that
there is a hard scattering component in virtual-photon proton interactions
leading to LRG events. A comparison of the energy flow in events with and
without a large rapidity gap showed that in LRG events the QCD radiative
processes are suppressed. 

The diffractive contribution to the proton structure function $F_2$ was
measured by H1~\cite{Hepdifff} and ZEUS~\cite{Zepdifff}. The diffractive
electron-proton cross section was found to be consistent with factorising
into a term describing the flux of a colourless component in the proton and
a term which describes the cross section for scattering of this colourless
object on an electron. LRG events were also observed in
photoproduction~\cite{Hgplrg,Zgplrgj}. A combined analysis of the
diffractive part of the proton structure function $F_2$ and the diffractive
photoproduction of jets indicated that a large fraction of the momentum of
the colourless object carried by partons is due to hard
gluons~\cite{Zepdiffg}. 

One of the most interesting questions raised by these LRG events is the
precise $W$ dependence of the cross section for diffractive scattering of
virtual photons on protons, $\gamma^* p \to Xp$. Here, $W$ is the $\gamma^*
p$ c.m. energy and the comparison should be done at fixed mass squared of
the virtual photon, $-Q^2$. In the Regge picture, the elastic and
diffractive cross sections in the forward direction are expected to behave
as (see e.g.~\cite{Collins}): 
\begin{eqnarray}
  d\sigma(t=0)/dt \propto
(W^2)^{2\alphapom(0)-2} \; \; \; , 
\label{eq:a} 
\end{eqnarray}
where $t$ is the square of the four-momentum transferred from the virtual
photon to the incoming proton. From elastic and total cross section
measurements for hadron-hadron scattering the intercept $\alphapom(0)$ of
the pomeron trajectory $\alphapom(t)$ was found to be 1.08~\cite{Donlan1}. A
similar energy dependence was observed for diffractive dissociation in
hadron-hadron scattering of the type $h_1 h_2 \rightarrow h_1 X $, for a
fixed mass $M_X$ of the diffractively produced system $X$ (see
e.g.~\cite{Goulian1,Cdf94}).  For DIS, with dominantly hard partonic
interactions, the BFKL formalism~\cite{Lipatov} leads to a pomeron intercept
of $\alphapom (0) \approx 1 + (12 \; \ln 2)\alpha_s/\pi \approx 1.5$ at $Q^2
= 20$~GeV$^2$ which could imply a rapid rise of the diffractive cross
section with $W$~\cite{Glr,Bartels}.

In the previous determinations of the diffractive structure function, the
subtraction of the nondiffractive contribution relied on specific
models~\cite{Hepdifff,Zepdifff}. In the present analysis the separation is
based on the data. The diffractive contr

ibution is extracted by a new method which uses the mass $M_X$ of the system
$X$, measured in the detector, to separate the diffractive and
nondiffractive contributions. The distribution in $\ln M^2_X$ exhibits, for
the nondiffractive component, an exponential fall-off towards small $\ln
M^2_X$ values, $dN^{nondiff}/d\ln M^2_X \propto \exp(b$~$\ln M^2_X)$, a
property which is predicted by QCD-based models for nondiffractive DIS (see
e.g.~\cite{ariadne,lund}). The parameter $b$ of this exponential fall-off is
determined from the data and is assumed to be valid in the region of overlap
between the diffractive and nondiffractive components, so allowing
subtraction of the nondiffractive background.

The cross section for diffractive production by virtual photons on protons,
$\gamma^*p \to XN$, is determined integrated over $t$. The system $N$ is
either a proton or a nucleonic system with mass $M_N < 4$~GeV. The 4~GeV
mass limit results from the acceptance of the detector. 

The prime goal of this analysis is the determination of the $W$ dependence
of the diffractive $\gamma^* p$ cross section in the range $60 < W <
245$~GeV, $M_X < 15$~GeV and $10 < Q^2 < 56$~GeV$^2$. The paper begins with
a brief introduction to the experimental setup and the event selection
procedure followed by a description of the determination of the mass $M_X$.
Using the measured $M_X$ distributions, the widely different behaviour of
the nondiffractive and diffractive contributions is demonstrated: production
of events with low masses $M_X$ is dominated by diffractive scattering while
nondiffractive events are concentrated at large $M_X$ values. These
observations lead to a straightforward procedure for extrapolating the
nondiffractive background into the low mass region and extracting the
diffractive contribution. An unfolding procedure is used to correct the
resulting number of diffractive events in $(M_X, W, Q^2)$ bins for detector
acceptance and migration effects. From the corrected number of events the
cross sections for diffractive production by virtual-photon proton
scattering are obtained and the $W$ dependence of diffractive scattering is
determined. Finally, the cross sections are analyzed in terms of the
diffractive structure function of the proton. 

\section{Experimental setup}

The experiment was performed at the electron-proton collider HERA using the
ZEUS detector. The analysis used data taken in 1993 where electrons of $E_e
= 26.7$~GeV collided with protons of $E_p = 820$~GeV. HERA is designed to
run with 210 bunches in each of the electron and proton rings.  In 1993, 84
paired bunches were filled for each beam and in addition 10 electron and 6
proton bunches were left unpaired for background studies. The integrated
luminosity was 543 nb$^{-1}$. Details on the operation of HERA and the
detector can be found in~\cite{Zepf2}. 

\subsection{ZEUS detector}

The analysis relies mainly on the high-resolution depleted-uranium
scintillator calorimeter and the central tracking detectors. The calorimeter
covers 99.7\% of the solid angle.  It is divided into three parts, forward
(FCAL) covering the pseudorapidity \footnote{The ZEUS coordinate system is
right-handed with the $Z$ axis pointing in the proton beam direction,
hereafter referred to as forward, and the $X$ axis horizontal, pointing
towards the center of HERA. The pseudorapidity $\eta$ is defined as $-\ln
(\tan \frac{\theta}{2})$, where the polar angle $\theta$ is taken with
respect to the proton beam direction from the nominal interaction point.}
region $4.3\geq \eta \geq 1.1$, barrel (BCAL) covering the central region
$1.1 \geq \eta \geq -0.75$ and rear (RCAL) covering the backward region
$-0.75\geq \eta \geq -3.8$. Holes of $20\times 20$~cm$^2$ in the center of
FCAL and RCAL are required to accommodate the HERA beam pipe.  The
calorimeter parts are subdivided into towers of typically $20 \times
20$~cm$^2$ transverse dimensions, which in turn are segmented in depth into
electromagnetic (EMC) and hadronic (HAC) sections.  To improve spatial
resolution, the electromagnetic sections are subdivided transversely into
cells of typically $5 \times 20$~cm$^2$ ($10 \times 20$~cm$^2$ for the rear
calorimeter).  Each cell is read out by two photomultiplier tubes, providing
redundancy and a position measurement within the cell. Under test beam
conditions \cite{Zcaltest}, the calorimeter has an energy resolution,
$\sigma_E $, given by $\sigma_E/E = 18\%/\sqrt{E}$ for electrons and
$\sigma_E/E = 35\%/\sqrt{E}$ for hadrons, where $E$ is in units of~GeV. In
addition, the calorimeter cells provide time measurements with a time
resolution below 1~ns for energy deposits greater than 4.5~GeV, a property
used in background rejection. The calorimeter noise, dominated by the
uranium radioactivity, in average is in the range 15-19~MeV for
electromagnetic cells and 24-30~MeV for hadronic cells. The calorimeter is
described in detail in~\cite{Zcaltest}. 

Charged particle detection is performed by two concentric cylindrical drift
chambers, the vertex detector (VXD) and the central tracking detector (CTD)
occupying the space between the beam pipe and the superconducting coil of
the magnet. The detector was operated with a magnetic field of 1.43 T. The
CTD consists of 72 cylindrical drift chamber layers organized into 9
superlayers~\cite{Foster}. In events with charged particle tracks, using the
combined data from both chambers, resolutions of $0.4$~cm in $Z$ and
$0.1$~cm in radial direction in the $XY$ plane are obtained for the primary
vertex reconstruction. From Gaussian fits to the $Z$ vertex distribution,
the rms spread is found to be $10.5$~cm in agreement with the expectation
from the proton bunch length. 

The luminosity is determined by measuring the rate of energetic
bremsstrahlung photons produced in the process $ep \to
ep\gamma$~\cite{lumi}. The photons are detected in a lead-scintillator
calorimeter placed at $Z=-107$~m. The background rate from collisions with
the residual gas in the beam pipe was subtracted using the unpaired electron
and proton bunches. 

\subsection{Kinematics}

The basic quantities used for the description of
inclusive deep inelastic scattering 
\begin{displaymath}
e(k)+p(P) \to e (k^\prime)+anything
\end{displaymath} 
are:   
\begin{eqnarray}
\qsd & = & -q^{2}  =  -(k - k^{\prime})^{2}  \; \; \; ,\\
  x &  = & \frac{\qsd}{2 P \cdot q} \; \; \; ,\\
  y &  = & \frac{ P \cdot q}{P \cdot k} \; \; \; ,\\
  W^2 & = &\frac{Q^2(1-x)}{x}+M_p^2 \approx \frac{\qsd}{x}   
        \;  {\rm for}\;  x \ll 1 \; \; \; ,
\end{eqnarray} 
where $k$ and $k^{\prime}$ are the four-momenta of the initial and
final state electrons, $P$ is the initial state proton four-momentum,
$M_p$ is the proton mass, $y$ is the fractional energy transfer to the
proton in its rest frame and $W$ is the $\gamma^* p$ c.m. energy. For
the range of $Q^2$ and $W$ considered in this paper we also have $W^2
\approx y\cdot s$, where $s=4 E_e E_p$ is the square of the $ep$
c.m. energy, $\sqrt{s} = 296$~GeV.

For the description of the diffractive processes, 
\begin{displaymath}
e  p \to e + X + N,
\end{displaymath}
in addition to the mass $M_X$, two further variables are introduced:
\begin{eqnarray}
\xpom & = & \frac{M^2_X + Q^2}{W^2 + Q^2} \; \; \; ,\\
\beta & = & \frac{Q^2}{M^2_X+Q^2} \; \; \; .
\end{eqnarray}

In models where diffraction is described by the exchange of a particle-like
pomeron, $\xpom$ is the momentum fraction of the pomeron in the proton and
$\beta$ is the momentum fraction of the struck quark within the pomeron. 

The kinematic variables $x$, $Q^2$ and $W$ were determined
with the double angle (DA) method~\cite{herakin}, in which only the angles of the
scattered electron 
($\theta^\prime_e$) and the hadronic system ($\gamma_H$) are used. This
reduces the sensitivity to energy scale uncertainties.  The
angle $\gamma_H$ characterizes the transverse and longitudinal
momenta of the hadronic system. In the na\"{\i}ve quark-parton model
$\gamma_H$ is the scattering angle of the struck quark. It was
determined from the hadronic energy flow measured in the
calorimeter. A momentum vector $\vec{p}_h \equiv (p_{{\rm X}},p_{{\rm Y}},
p_{{\rm Z}})$ was assigned to each cell $h$
with energy $E$ in such a way that $\vec{p}_h^2 = E^2$.  The cell angles
were calculated from the geometric center of the cell and the vertex
position of the event. The angle $\gamma_H$ was calculated according
to
\begin{eqnarray}
\cos \gamma_H = \frac{(\sum_{h} p_{{\rm X}})^{2}+(\sum_{h} p_{{\rm Y}})^{2} 
- (\sum_{h} ( E-p_{{\rm Z}}))^{2}}
{(\sum_{h} p_{{\rm X}})^{2}+(\sum_{h} p_{{\rm Y}})^{2} + (\sum_{h} 
( E-p_{{\rm Z}}))^{2}}, 
\label{cosg}
\end{eqnarray}
where the sums, $\sum_{h}$, here and in the following, run over all
calorimeter cells $h$ which were not assigned to the scattered electron. The
cells were required to have energy deposits above 60~MeV in the EMC section
and 110~MeV in the HAC section and to have energy deposits above 140~MeV
(160~MeV) in the EMC (HAC) sections, if these energy deposits were isolated.
The last two cuts remove noise caused by the uranium radioactivity which
affects the reconstruction of the DA variables at low $M_X$. 

In the double angle method, in
order that the hadronic system be well measured, it is necessary to
require a minimum of hadronic activity in the calorimeter away from the
forward direction. A suitable quantity for this purpose is the
hadronic estimator of the variable $y$~\cite{Jacquet}, defined by
\begin{eqnarray}
\yjb = \frac{\sum_{h} \left( E-p_{{\rm Z}}\right) }{2E_{e}}.
\label{yjabl}
\end{eqnarray}

We study below events of the type
\begin{displaymath} 
ep \to e + X  +  rest,
\end{displaymath}
where $X$ denotes the hadronic system observed in the detector
and $rest$ the particle system
escaping detection through the beam holes.  The mass $M_X$ of the system
$X$ is determined from the energy deposited in the CAL cells
according to:
\begin{eqnarray} 
(M_X^{meas})^2 = (\sum_{h} E)^{2}-(\sum_{h} p_{X})^{2}-(\sum_{h}
p_{Y})^{2}-(\sum_{h} p_{Z})^{2} \; \; \; .
\label{mass} 
\end{eqnarray}

\subsection{Event selection}

The event selection at the trigger level was identical to that used for our
$F_2$ analysis~\cite{Zepf2}. The off-line cuts were very similar to those
applied in the double angle analysis of $F_2$~\cite{Zepf2}. The resulting
event sample is also almost identical to the one used for our recent studies
of large rapidity gap events in DIS~\cite{Zeplrgj,Zeplrgf}. For ease of
reference we list the main kinematic requirements imposed, which limit the
$W$ and $Q^2$ range of the measurement: 
\begin{itemize}
\item $E^\prime_e >$~8~GeV, where $E^\prime_e$ is the energy of the
scattered electron, to have reliable electron finding and to control the
photoproduction background; 
\item $y_e<0.95$, where $y_e$ is the variable $y$ calculated from the
scattered electron, to reject spurious low energy electrons, especially in
the forward direction,
\item the impact point of the electron on the face of the RCAL had to
lie outside a square of side 32~cm centered on the beam axis (``box cut''),
to ensure full containment of the electron shower,
\item $y_{JB}>0.02$, to ensure a good measurement of the angle $\gamma_H$
and of $x$,
\item $35 < \delta < 60$~GeV, where $\delta = \sum_{h}
\left(E-p_{Z}\right)$, to control radiative corrections and reduce
photoproduction background. 
\end{itemize}

The differences with respect to the event selection used for the $F_2$
analysis in ~\cite{Zepf2} are an increase of the lower limit on $E^\prime_e$
from 5 to 8~GeV and a lowering of the $y_{JB}$ cut from 0.04 to 0.02 which
became possible with the improved noise suppression procedure explained
above. The increase of the $E^\prime_e$ limit reduces background from
photoproduction; the lower $y_{JB}$ cut extends the acceptance towards lower
$W$ values. It was checked that the $F_2$ values obtained with the modified
noise procedure were fully compatible with the values published
previously~\cite{Zepf2} in the whole $Q^2$, $W$ range investigated in the
present paper. 

The primary event vertex was determined from tracks reconstructed using
VXD+CTD information. If no tracking information was present the vertex
position was set to the nominal interaction point. 

After the selection cuts and the removal of QED Compton scattering events
and residual cosmic-ray events, the DIS sample contained 46k events. For the
analysis of diffractive scattering, events with $Q^2 > 10$~GeV$^2$ were
used. The background from beam gas scattering in this sample was less than
1\% as found from the data taken with unpaired bunches. 

\section{Simulation and method of analysis}

\subsection{Monte Carlo simulation}

Monte Carlo simulations were used for unfolding the produced event
distributions from the measured ones, for determining the acceptance and for
estimating systematic uncertainties. 
 
Events from standard DIS processes with first order electroweak corrections
were generated with HERACLES 4.4~\cite{herac}. It was interfaced using
Django 6.0~\cite{django} to ARIADNE 4.03~\cite{ariadne} for modelling the
QCD cascade according to the version of the colour dipole model that
includes the boson-gluon fusion diagram, denoted by CDMBGF.  The
fragmentation into hadrons was performed with the Lund fragmentation scheme
\cite{lund} as implemented in JETSET 7.2 \cite{jetset}. The parton densities
of the proton were chosen to be the MRSD$^\prime$- set~\cite{Martin}. Note
that this Monte Carlo code does not contain contributions from diffractive
$\gamma^*p$ interactions. 

In order to model the DIS hadronic final states from diffractive
interactions where the proton does not dissociate,
\begin{displaymath} 
 ep \to e + X + p,
\end{displaymath} 
two Monte Carlo event samples were studied, one of which was generated by
POMPYT 1.0~\cite{pompyt}. POMPYT is a Monte Carlo realization of
factorizable models for high energy diffractive processes where, within the
PYTHIA 5.6~\cite{pythia} framework, the beam proton emits a pomeron, whose
constituents take part in a hard scattering process with the virtual-photon.
For the quark momentum density in the pomeron it has been common to use the
so-called Hard POMPYT version, $\beta f(\beta) \propto \beta \cdot (1 -
\beta)$. For this analysis the form
\begin{eqnarray}
\beta f(\beta) = constant \cdot \beta
\end{eqnarray}
was used which enhances the rate of low $M_X$ events as preferred by the data.

The second sample was generated following the Nikolaev-Zakharov (NZ) model
\cite{Nz} which was interfaced to the Lund fragmentation scheme
\cite{Solano}. In the NZ model, the exchanged virtual photon fluctuates into
a $q\bar{q}$ pair or a $q\bar{q}g$ state which interacts with a colourless
two-gluon system emitted by the incident proton. In the Monte Carlo
implementation of this model the mass spectrum contains both components but
the $q\bar{q}g$ states are fragmented into hadrons as if they were a
$q\bar{q}$ system with the same mass $M_X$. Hadronic final states $X$ are
generated only with masses $M_X>$~1.7~GeV. For a description of the NZ model
see also~\cite{Zepdifff}. 

All Monte Carlo events were passed through the standard ZEUS detector and
trigger simulations and the event reconstruction package. 

\subsection{Weighting of diffractive Monte Carlo events}
\label{s:Weighting}

In order to determine from the number of observed events the number of
produced events in each bin an unfolding procedure based on a weighted Monte
Carlo sample was applied. The unfolding procedure is most reliable if the
Monte Carlo event distributions are in agreement with the data. POMPYT was
used for unfolding. However, POMPYT as well as the NZ model showed
considerable discrepancies relative to the measured $W$ distributions in the
kinematic range of this study. This problem was overcome as follows:  to
account for the lack of diffractive events in the low mass region, $M_X <
1$~GeV, events were generated separately for $\rho^o$ production via
$\gamma^* p \to \rho^o p$~\cite{DIPSI} and added to the POMPYT event sample.
The number of $\rho^o$ events and their distribution as a function of $W$
and $Q^2$ were determined from the analysis of this
experiment~\cite{Zeprho}. Furthermore, the POMPYT and $\rho^o$ events were
weighted to agree with a Triple Regge~\cite{Mueller,Fiefox} inspired model
(TRM) predicting for the diffractive cross section:
 
\begin{eqnarray}
\frac {d\sigma^{diff}_{\gamma^*p \to XN}(M_X,W,Q^2)}{dM_X } & = &
  C\cdot (1+c_L\cdot Q^2/M_X^2)\, (M_0^2+Q^2)^{-\alpha_k(0)} 
   (M_X^2/(M_X^2+Q^2)) \nonumber \\ 
 & &\cdot\frac {M_X}{(M_X^2+Q^2)^{2-\alpha_k(0)}} 
    \cdot (\frac{W^2}{M_X^2+Q^2})^{2\overline{\alphapom} -2} .
\label{eq:TRMweight}
\end{eqnarray}

Here $C$ is a normalization constant, $\alpha_k(0) = 0$ ($\alpha_k(0) = 1$)
for $M_X < M_0$ ($M_X \ge M_0$) and $\overline{\alphapom}$ is the pomeron
trajectory averaged over the square of the four-momentum transfer, $t$,
between the incoming and the outgoing proton. The parameters $c_L$, $M_0$,
$\overline{\alphapom}$ of the TRM model were determined in the unfolding
(see section~\ref{s:detercross} below) and will be referred to as
``weighting parameters''. The weighted sample of POMPYT events will be
referred to as ``weighted POMPYT''.

\subsection{Mass determination}

The mass of the system $X$ was determined from the energy deposits in the
calorimeter using Eq.~\ref{mass}. The mass $M_X^{meas}$ measured in this way
has to be corrected for energy losses in the inactive material in front of
the calorimeter and for acceptance. The correction was determined by
comparing for Monte Carlo (MC) generated events the MC measured mass,
$M_X^{MCmeas}$, to the generated mass, $M_X^{MCgen}$, of the system $X$. The
mass correction was performed in two steps. In the first step an overall
mass correction factor was determined. In the second step the diffractive
cross sections were determined by an unfolding procedure (see
section~\ref{s:detercross}) taking into account for each ($M_X, W, Q^2$)
interval the proper mass correction as determined from the MC simulation. 

The overall correction factor $f(M_X)$ was determined from the average ratio
of measured to generated mass $M_X$,
\begin{displaymath}
\label{masscorr} 
 f(M_X^{MCmeas}) = \frac {M_X^{MCmeas}}{M_X^{MCgen}} \; \; \; ,
\end{displaymath}
as a function of $M_X$, $W$ and $Q^2$. The dependence of $f(M_X^{MCmeas})$
on $M_X, W, Q^2$ was found to be sufficiently small ($\pm 6\%$) for $M_X >
1.5$~GeV so that it could be neglected in the first step of the mass
correction. The average correction factor was $f(M_X^{MCmeas})=0.68$.  The
same correction factor was used for masses below 1.5~GeV.  The correction
factor $f = 0.68$ was applied to obtain from the measured mass the corrected
mass value, $M_X^{cor}=M_X^{meas}/f$. 

\begin{figure}[ht]
\includegraphics{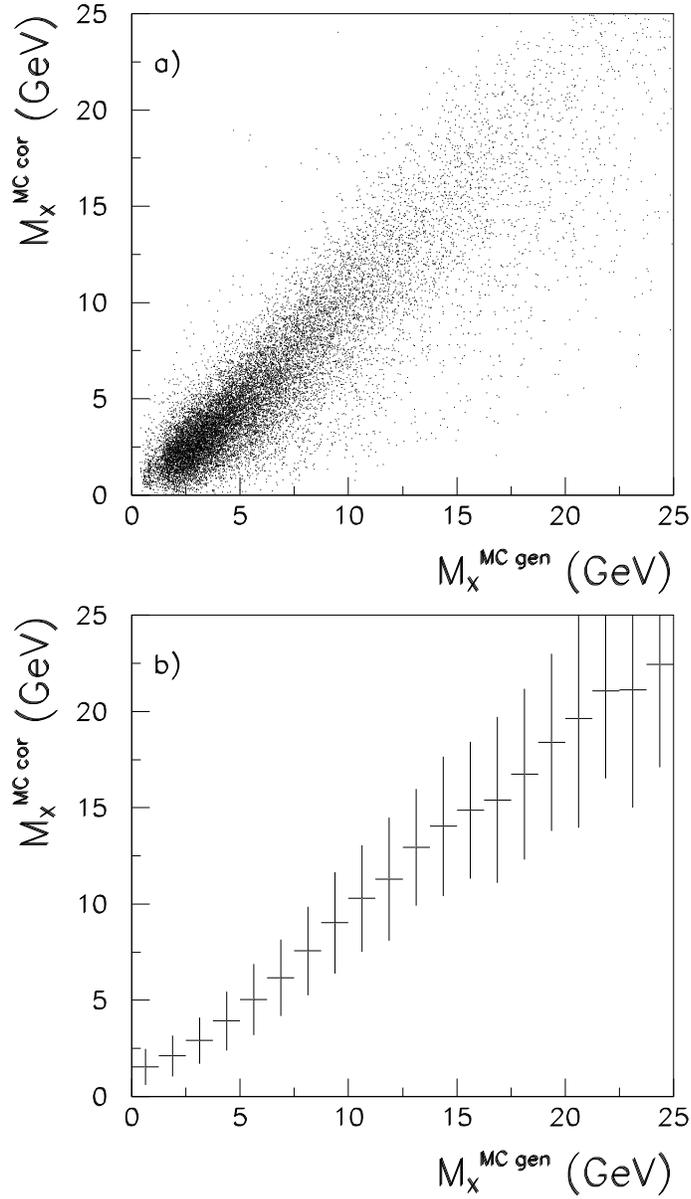}
\unitlength1cm
\begin{picture}(15,20)
\thicklines
\end{picture}
\caption{\label{f:mxcor}
        { 
             a) The corrected mass $M^{MCcor}_X$ versus the generated mass
             $M^{MCgen}_X$ for MC events generated with POMPYT in the $W, Q^2$
             range of this measurement. 
             b) The same for the average values of $M^{MCcor}_X$ and
             $M^{MCgen}_X$. The vertical bars show the rms uncertainty
             of $M^{MCcor}_X$ for a single measurement.}}
\end{figure}

Figures~\ref{f:mxcor}a,b show, for MC events, the corrected versus the
generated $M_X$. The error bars in Fig.~\ref{f:mxcor}b give the rms
resolution for a single $M_X$ measurement. A tight correlation between
corrected and generated mass is observed except when $M_X < 2$~GeV where the
mass resolution is comparable to the value of the mass. The mass resolution
increases smoothly from $1$~GeV near the $\rho$ mass to $1.3$~GeV
($2.9$~GeV) at $M_X = 3$~GeV ($15$~GeV). For $M_X^{MCgen}>3$~GeV it can be
approximated by $\sigma (M_X^{MCcor})/\sqrt {M_X^{MCgen}}=0.75$~GeV$^{1/2}$. 

A test of the MC predictions for the mass measurement at low $M_X$ values
was performed by studying the reaction
\begin{displaymath} 
ep \to e + \rho^o + p  \; \; \; ,
\end{displaymath}
where the pions from the decay $\rho^o \to \pi^+ \pi^-$ were measured with
the central tracking detector~\cite{Zeprho}. The $\pi^+\pi^-$ mass
resolution from tracking was 25~MeV(rms). From a total of 60 events with
$700<M_{\rho^o}^{tracking}<800$~MeV in the kinematic range $Q^2 = 7 -
25$~GeV$^2$, $W = 60 - 134$~GeV, an average $M_X^{cor}$ of $1.2$~GeV and an
average mass resolution $\sigma(M_X^{cor})$ of 0.9~GeV were obtained; all
but 4 events were reconstructed with a mass $M^{cor}_X$ below 3.0~GeV.  The
Monte Carlo simulation for this channel predicted $M_X^{MCcor} = 1.2$~GeV
and $\sigma(M_X^{MCcor}) = 0.8$~GeV, in good agreement with the data. 

All $M_X$ results presented below refer to $M_X^{cor}$.

\subsection{Acceptance for diffractive events}

\begin{figure}[ht]
\includegraphics{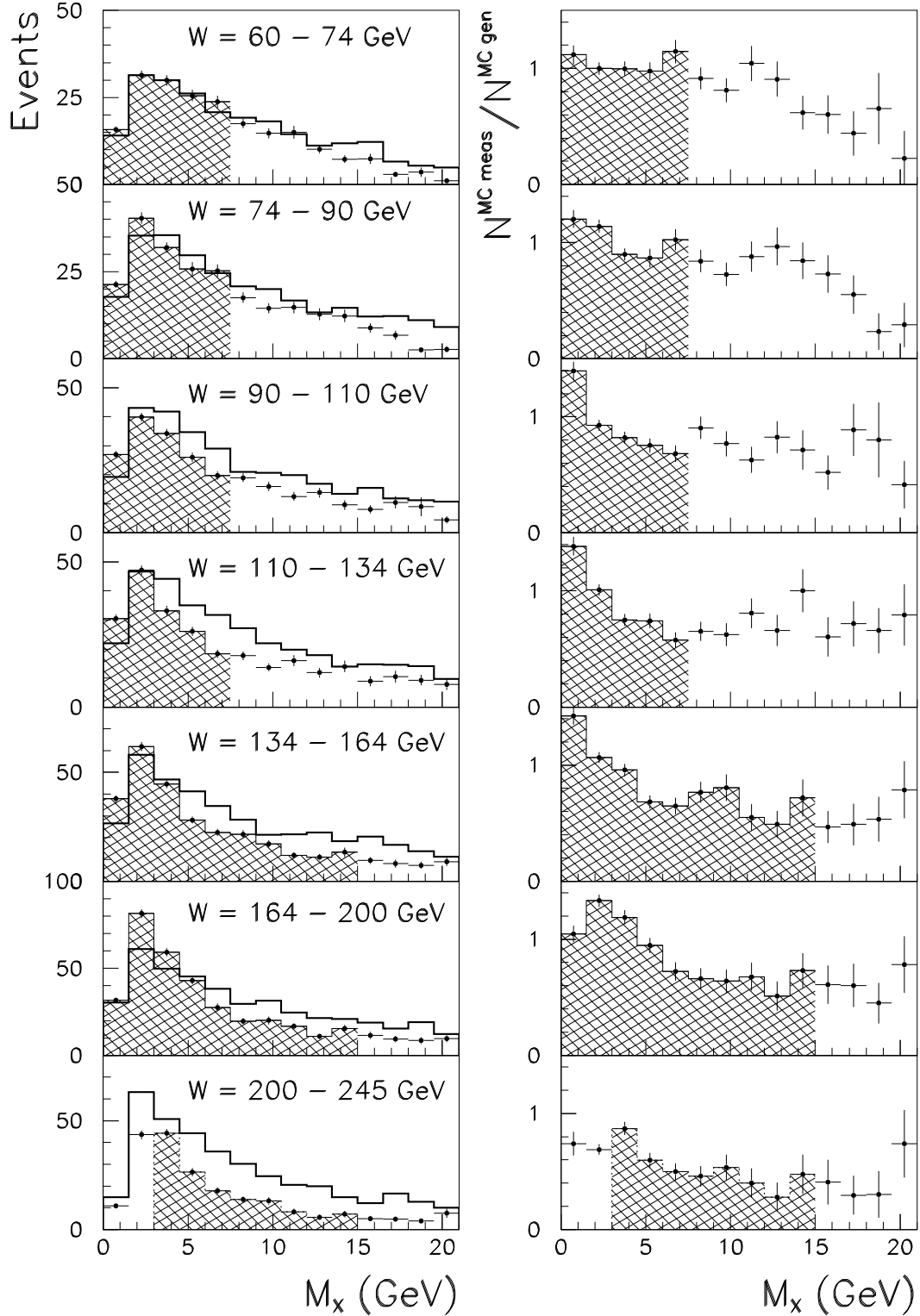}
\unitlength1cm
\begin{picture}(15,20)
\thicklines
\end{picture}
\caption{\label{f:nmeasngen}
        {    Left: $M_X$ distributions for MC events on the generator level
             (histograms) and after reconstruction and measurement simulation               (points with error bars)
             as a function of $M_X$ for different $W$ bins at
             $Q^2 = 14$~GeV$^2$ for weighted POMPYT (see text). Right: The 
             ratio ${\cal N}^{MCmeas}/{\cal N}^{MCgen}$ of measured and
             generated MC event numbers as a function
             of $M_X$. In all plots the hashed areas show the $M_X$ regions 
             used for the determination of the diffractive cross section.}}
\end{figure}

\begin{figure}[ht]
\includegraphics{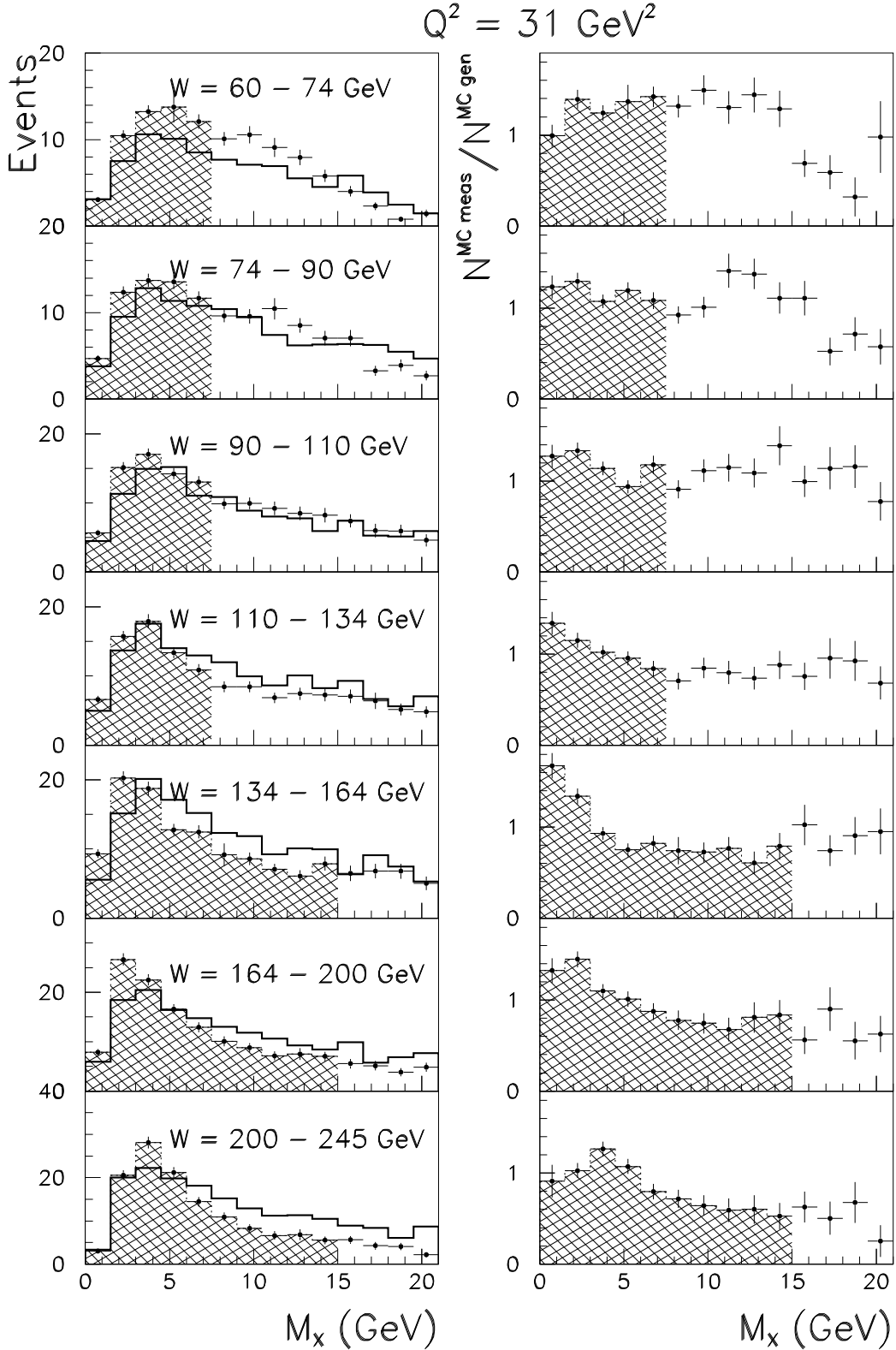}
\unitlength1cm
\begin{picture}(15,20)
\thicklines
\end{picture}
\caption{\label{f:nmeasngen2}
        {    Left: $M_X$ distributions for MC events on the generator level
             (histograms) and after reconstruction and measurement simulation               (points with error bars)
             as a function of $M_X$ for different $W$ bins at
             $Q^2 = $31~GeV$^2$ for weighted POMPYT (see text).
             Right: The ratio ${\cal N}^{MCmeas}/{\cal N}^{MCgen}$ of 
             measured and generated MC event numbers as a 
             function of $M_X$. In all plots the dashed areas show the $M_X$ 
             regions used for the determination of the diffractive cross 
             section.}}
\end{figure}

A measure of the acceptance for diffractive events is the ratio ${\cal R} =
{\cal N}^{MCmeas}/{\cal N}^{MCgen}$ of events measured to events generated
in an ($M_X, W, Q^2$) bin using $M^{cor}_X$. Figures~\ref{f:nmeasngen}
and~\ref{f:nmeasngen2} show, for weighted POMPYT events, the distributions
of ${\cal N}^{MCgen}$ (histograms) and ${\cal N}^{MCmeas}$ (solid points)
for the ($W$, $Q^2$) bins used in this analysis (see
section~\ref{s:binning}). Here, the generated values for $M_X, W, Q^2$ were
used for $ {\cal N}^{MCgen}$ while $M_X^{cor}$ and the double-angle
quantities for $W$ and $Q^2$ were used for ${\cal N}^{MCmeas}$, as in the
analysis of the data. The $M_X$ distributions increase from small $M_X$
values to a maximum at $M_X = 2 - 5$~GeV and then fall off towards higher
masses. There is some leakage of events into the low $M_X$ bin as seen in
the ratio $\cal R$ shown in the right-hand parts of
Figs.~\ref{f:nmeasngen},~\ref{f:nmeasngen2}. 

The shaded areas mark the $M_X$ regions used for extraction of the
diffractive cross sections. The ratio $\cal R$ is above unity at small $M_X$
as a result of the migration from higher $M_X$ masses; for larger $M_X$
values $\cal R$ is rather constant and between $70$ and $100\%$ in the bins
considered for the analysis, except in the highest $W$ interval for $Q^2 =
14$~GeV$^2$ where the acceptance is around $80\%$ at low masses falling to
about $50\%$ at $M_X = 15$~GeV. This is caused by the reduced efficiency for
detecting the scattered electron and by the requirement that $\delta =
\sum_{h}(E - p_Z) > 35$~GeV. 

\subsection{ General characteristics of the $M_X$ distributions }

\begin{figure}[ht]
\includegraphics{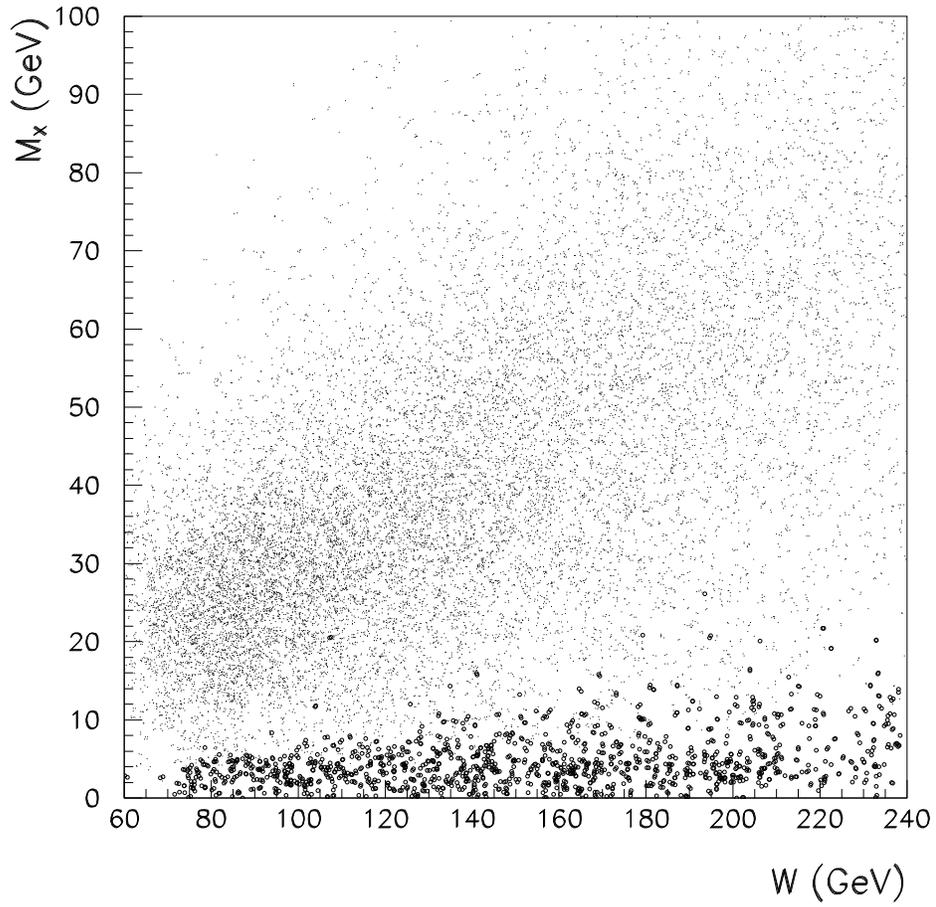}
\unitlength1cm
\begin{picture}(15,20)
\thicklines
\end{picture}
\caption{\label{f:mxvsw}
        {    DIS events from data measured in the $Q^2$ range 
             10 - 56~GeV$^2$: the scatter plot of $M_X$ versus $W$. 
             The events with $\eta_{max} < $1.5 are shown as larger dots; they 
             concentrate at small values of $M_X$.}}
\end{figure}

The method of separating the diffractive and nondiffractive contributions is
based on their very different $M_X$ distributions. As a first illustration,
Fig.~\ref{f:mxvsw} shows the distribution of $M_X$ versus $W$ for the data.
Two distinct classes of events are observed, one concentrated at small
$M_X$, the second extending to large values of $M_X$. Most of the events in
the low $M_X$ region exhibit a large rapidity gap, which is characteristic
of diffractive production. This is shown by Fig.~\ref{f:mxvsw} where the
events with a large (small) rapidity gap, $\eta_{max} < 1.5$ ( $\eta_{max} >
1.5$) are marked by different symbols. Here $\eta_{max}$ is the
pseudorapidity of the most forward going particle. For this analysis a
particle is defined as an isolated set of adjacent calorimeter cells with
more than 400~MeV summed energy, or a track observed in the central track
detector with more than 400~MeV momentum. A cut of $\eta_{max} < 1.5$
corresponds to a visible rapidity gap larger than 2.2 units since no
particles were observed between the forward edge of the calorimeter ($\eta =
3.7 - 4.3$) and $\eta = 1.5$. For $\eta_{max} < 1.5$ the contribution from
nondiffractive scattering is expected to be
negligible~\cite{Zeplrg,Zepdifff}.

\begin{figure}[ht]
\includegraphics{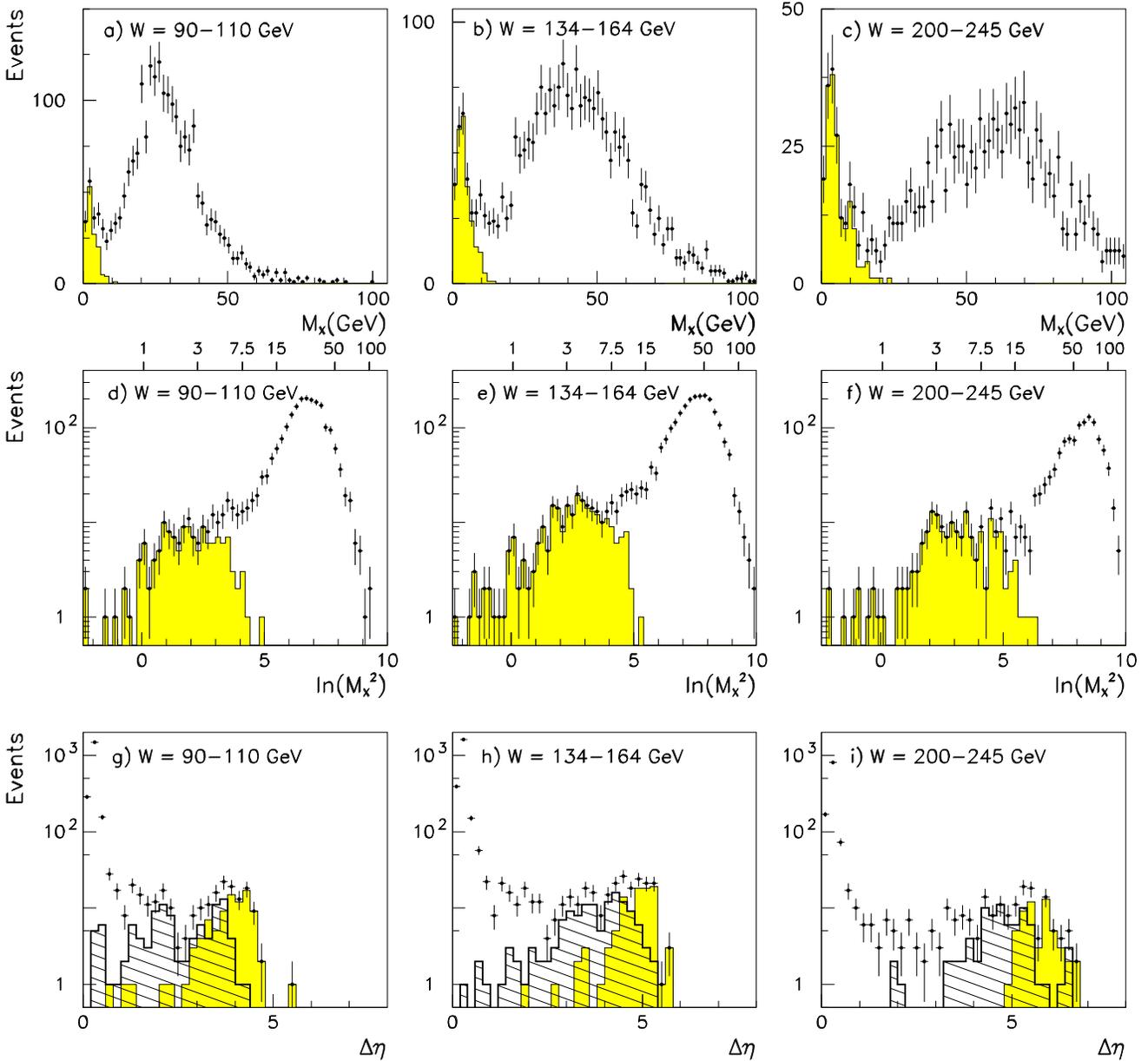}
\unitlength1cm
\begin{picture}(15,20)
\thicklines
\end{picture}
\caption{\label{f:madist}
        { Distributions of $M_X$ at $Q^2=$ 14~GeV$^2$: a) - c) for the $W$ 
          intervals 90 - 110,  134 - 164,  200 - 245~GeV.
          d) - f) Distributions of $\ln M^2_X$ for the same $W$ intervals.
          The shaded histograms in a) - f) show the events which have 
          $\eta_{max} < 1.5$ corresponding to a rapidity gap in the detector 
          larger than 2.2 units. 
          g) - i) Distributions of the rapidity gap $\Delta\eta$ for the same
          $Q^2$ and $W$ values.  The histograms show the distributions for
          events with $M_X <$~3~GeV (shaded) and $M_X =$~3-7.5~GeV 
         (skewed hatched). Here $M_X$ is the corrected mass.
          The distributions are not corrected for 
          acceptance effects.}}
\end{figure}

The measured $M_X$ distributions are shown in Figs.~\ref{f:madist}(a-c) for
three $W$ intervals, $W = 90 - 110, 134 - 164$ and $200 - 245$~GeV at $Q^2$
= 14~GeV$^2$. The distributions are not corrected for acceptance. For all
$W$ bins two distinct groups of events are observed, one peaking at low
$M_X$ values, the other at high $M_X$ values. While the position of the low
mass peak is independent of $W$, the high mass peak moves to higher values
as $W$ increases. As already seen, most events in the low mass peak possess
a large rapidity gap. This is illustrated by the shaded histograms which
represent the events with $\eta_{max} < 1.5$.

The size of the rapidity gap, $\Delta \eta$, can be seen from
Figs.~\ref{f:madist}(g - i) which show the distributions of the rapidity gap
between the edge of the calorimeter ($\eta \approx 3.9$, which is the $\eta$
value of the geometric center of the HAC cells closest to the proton beam)
and the most forward lying cell with energy deposition greater than 200~MeV
(the threshold is reduced in comparison with the determination of
$\eta_{max}$ because here single cells are considered instead of a group of
cells).  The plots give the distribution of all events (points with error
bars) and those with $M_X <$~3~GeV (shaded) and 3 - 7.5~GeV (skewed
hatching).  There is a strong concentration of events at small rapidities,
$\Delta\eta$ $<1$, which stem from non diffractive processes.  The
distributions demonstrate that the majority of low $M_X$ events are
associated with a large rapidity gap $\Delta\eta~>2$.  The average rapidity
gap increases with growing $W$. For a given $W$ value $\eta_{max}$ is
correlated with the maximum possible $M_X$ value but does not allow the
determination of $M_X$ uniquely because of fragmentation effects.  Since our
aim is the determination of the diffractive cross section as a function of
$M_X$, the analysis was based on $M_X$ and not on $\eta_{max}$.

\begin{figure}[ht]
\includegraphics{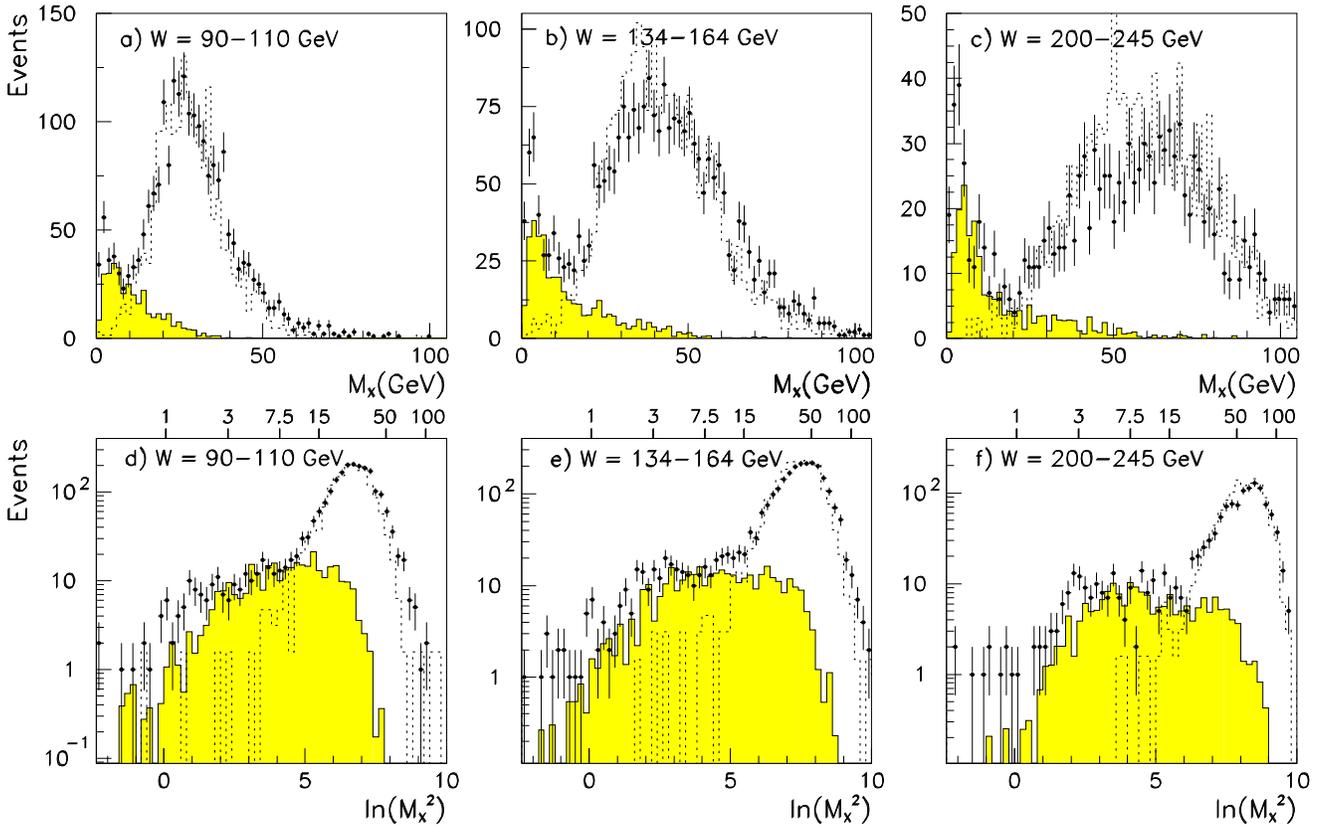}
\unitlength1cm
\begin{picture}(15,20)
\thicklines
\end{picture}
\caption{\label{f:madist2}
        { Distributions of $M_X$ at $Q^2=$14~GeV$^2$: a) - c) for the $W$ 
          intervals 90 - 110,  134 - 164,  200 - 245~GeV, d) - f)
          plotted versus $\ln M^2_X$ for the same $W$ intervals as in a) - c). 
          Here $M_X$ is the corrected mass. The distributions are 
          not corrected for acceptance effects.
          Shaded histograms show the prediction of the NZ model for 
          diffractive production; dashed histograms show the prediction of 
          CDMBGF for nondiffractive production. 
          The MC events were passed through the standard ZEUS detector 
          simulation. The CDMBGF distributions were normalized 
          to 85$\%$ of the data while the NZ distributions were 
          normalized to the observed number of diffractive events with
          $M_X >$~1.7~GeV; in the NZ model, diffractive events are generated
          for $M_X >$~1.7~GeV.}}
\end{figure}

In Fig.~\ref{f:madist2} the measured $M_X$ distributions are compared with
the NZ and CDMBGF predictions. The shaded distributions show the NZ
predictions for diffractive production. They peak at small masses. The
predictions of CDMBGF for nondiffractive production (dotted histograms) peak
at high masses. The sum of the NZ and CDMBGF contributions reproduce the
main features of the data which are the low and high mass peaks. 

The properties of the $M_X$ distributions can be understood best when
plotted as a function of $\ln M_X^2$, shown in Figs.~\ref{f:madist}(d-f) and
Figs.~\ref{f:madist2}(d-f).  Here, and in the following, masses and energies
are given in units of~GeV. In this representation the low mass peak shows up
as a plateau-like structure at low $\ln M_X^2$, most notably at high $W$
values. The high mass peak exhibits a steep exponential fall-off towards
smaller $\ln M_X^2$ values.  The shape of the exponential fall-off is
independent of $W$, a property which is best seen when the $\ln M_X^2$
distributions are replotted in terms of the scaled variable [$\ln M_X^2$ +
$\ln (s/W^2)$] (the total $ep$ c.m. energy squared, $s$, is introduced for
convenience). This is shown in Fig.~\ref{f:scalmadist} at $Q^2 = 14$ and
31~GeV$^2$ where the scaled $\ln M_X^2$ distributions are overlaid for three
$W$ intervals. The position of the high mass peak in $\ln M_X^2$ grows
proportionally to $\ln W^2$ and the slope of the exponential fall-off
towards small $\ln M_X^2$ values is approximately independent of $W$.

\subsection{$M_X$ dependence of the nondiffractive contribution}
\label{s:lpsnondiff}

\begin{figure}[ht]
\includegraphics{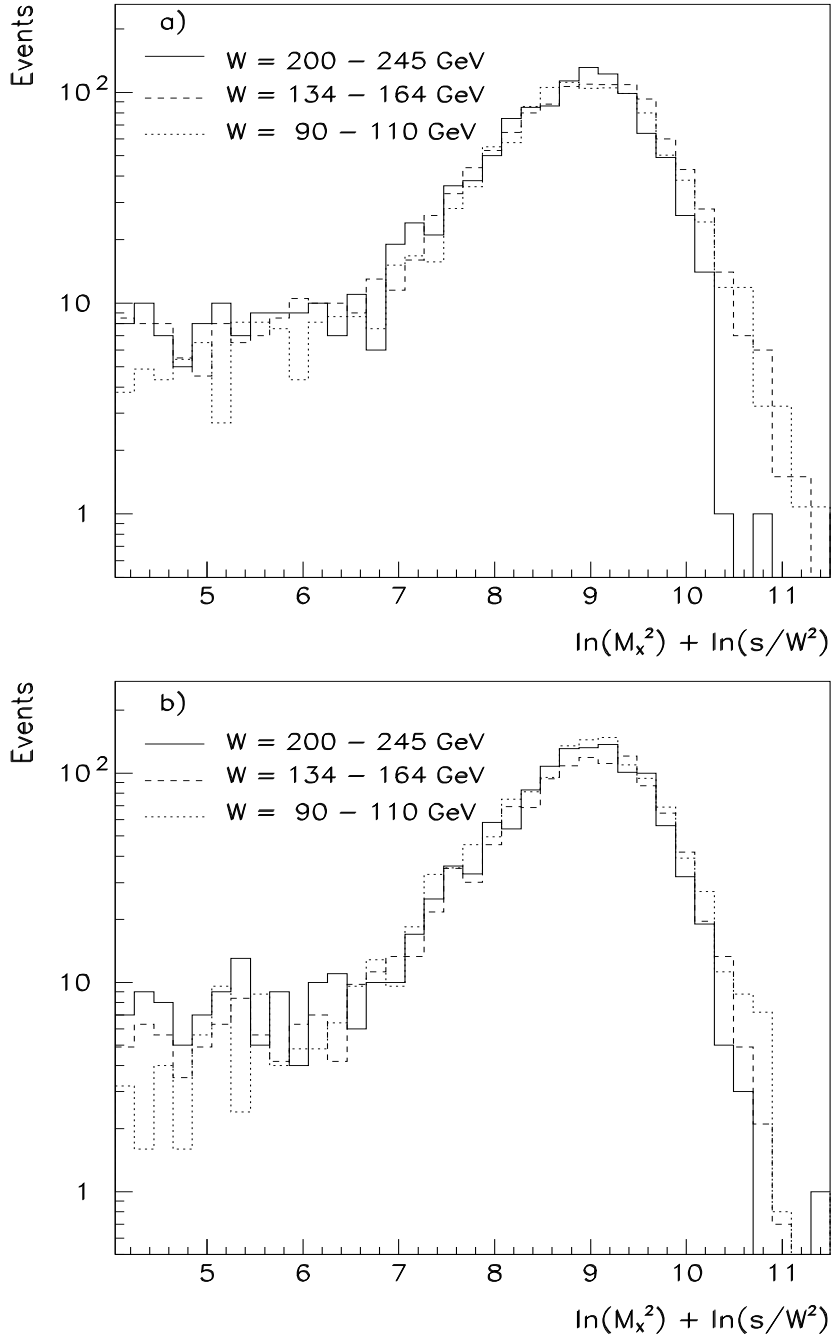}
\unitlength1cm
\begin{picture}(15,20)
\thicklines
\end{picture}
\caption{\label{f:scalmadist}
        {    Distributions of $\ln M^2_X$ + $\ln (s/W^2)$ for the $W$ 
             intervals 90 - 110~GeV (dotted),  134 - 164~GeV (dashed), 
             200 - 245~GeV (solid) 
             ($\ln W^2$ = 9.0 - 9.4, 9.8 -10.2, 10.6 - 11.0)
             at a) $Q^2 = $14~GeV$^2$ and b) 31~GeV$^2$. 
             Here $M_X$ is the corrected mass;
             the distributions are the measured ones, not corrected for
             acceptance effects. For each $Q^2$ the three distributions were 
             normalized to the same number of events.}}
\end{figure}

While in diffractive scattering the outgoing nucleonic system remains
colourless, in nondiffractive DIS the incident proton is broken up and the
remnant of the proton is a coloured object. This gives rise to a substantial
amount of initial and final state QCD radiation, followed by fragmentation,
between the directions of the incident proton and the current jet as
illustrated in Fig.~\ref{f:disdiaga}. The salient features of the resulting
$M_X$ distribution, namely the exponential fall-off and the scaling in [$\ln
M_X^2$ + $\ln (s/W^2)$], can be understood from the assumption of uniform,
uncorrelated particle emission in rapidity ${\cal Y}$ along the beam axis in
the $\gamma^*p$ system~\cite{feyn}: 
\begin{eqnarray}
  dN_{part}/d{\cal Y} = \lambda, \; \; \; \; \lambda = \rm{constant.}
\label{eq:rapdens}
\end{eqnarray}
At the $Q^2$ values under study, for ${\cal Y} > 0$, the beam axis in the
$\gamma^*p$ system is approximately given by the proton direction in the
HERA system. Since the shape of the rapidity distribution is invariant under
translations along the $\gamma^*$ beam axis we translate the ${\cal Y}$
distribution measured in the $\gamma^* p$ system until the point of maximum
rapidity agrees with the maximum rapidity ${\cal Y}_{max}$ in the $ep$
system.  For an (idealized) uniform ${\cal Y}$ distribution between maximum
and minimum rapidities of ${\cal Y}_{max}$ and ${\cal Y}_{min}$,
respectively, the total center of mass energy $W$ is given by
\begin{eqnarray} 
W^2 =  c_0\cdot \exp ({\cal Y}_{max} - {\cal Y}_{min}), \; \; \; {\rm{assuming}} \; \; \; ({\cal Y}_{max} - {\cal Y}_{min}) \gg 1 .
\end{eqnarray}
Here, $c_0$ is a constant.
The mass $M_X$ of the particle system that can be observed in the detector
is reduced by particle loss mainly through the forward beam hole:
\begin{eqnarray} 
M_X^2 = c_0\cdot \exp ({\cal Y}^{det}_{limit} - {\cal Y}_{min}) = W^2 \cdot
\exp ({\cal Y}^{det}_{limit} - {\cal Y}_{max})
\label{eq:ylimit}
\end{eqnarray}
where ${\cal Y}^{det}_{limit}$ denotes the limit of the FCAL acceptance
(neglecting the mass and transverse momentum of the produced particles).
Equation~\ref{eq:ylimit} predicts scaling of the $\ln M^2_X$ distribution
when plotted as a function of ~$\ln (M^2_X/W^2)$, in agreement with the
behaviour of the data seen in Fig.~\ref{f:scalmadist} where the data are
plotted in terms of [$\ln M^2_X + \ln (s/W^2)$]. The quantity (${\cal
Y}_{max} - {\cal Y}^{det}_{limit}$) is the effective width of the beam hole
and can be estimated from the effective maximum rapidity and the detector
geometry. 

The mass $M_X$ is expected to fluctuate statistically due to a finite
probability $P(0)$ that no particles are emitted between ${\cal
Y}^{det}_{limit}$ and ${\cal Y}^{det}_{limit} - \Delta {\cal Y}$. This
generates a gap of size $\Delta {\cal Y}$. The assumption of uncorrelated
particle emission leads to Poisson statistics which predicts $P(0) =
exp(-\lambda\Delta {\cal Y})$ resulting in an exponential fall-off of the
$\ln M^2_X$ distribution, \begin{eqnarray} \frac{d{\cal N}^{nondiff}}{d\ln
M^2_X} = c\, \exp (b \ln M^2_X) \; \; \; , \label{lmdist} \end{eqnarray}
where the slope $b$ is equal to the parameter $\lambda$ and $c$ is a
constant. The exponential fall-off of the $\ln M_X^2$ distribution towards
small values of $\ln M^2_X$ expected from this simplified consideration is
indeed observed in models which include QCD leading order matrix elements,
parton showers and fragmentation such as CDMBGF shown by the dashed
histograms in Figs.~\ref{f:madist2}(d-f).  This shows that the known sources
of long range correlations like conservation of energy and momentum, of
charge and of colour, which are incorporated in CDMBGF, do not lead to
significant deviations from an exponential behaviour with the possible
exception of a very small fraction ($0.2 - 0.4\%$) of the CDMBGF events
which is found above the exponential at low $\ln M^2_X$ values. 

In principle, the exponential fall-off of the $\ln M_X^2$ distribution
should start at the maximum value of $\ln M_X^2$ allowed by kinematics and
acceptance, Max$(\ln M_X^2) = \ln W^2 - (2$ to $3)$. The data in
Fig.~\ref{f:madist}d-f and Fig.~\ref{f:scalmadist} break away from the
exponential behaviour towards high values of $\ln M_X^2$ leading to a
rounding-off. It mainly results from the finite size of the selected $W$
intervals, the edge of the calorimeter acceptance in the forward direction
($\eta_{edge}$ = 3.7 to 4.3) and the finite resolutions with which $W$ and
$M_X$ are measured. With good accuracy the exponential fall-off is observed
for $\ln M_X^2 \leq \ln W^2 -\eta_0$, with $\eta_0 \approx 3.0$, over more
than two units of rapidity (see also Section 4.2). 

\begin{figure}[ht]
\includegraphics{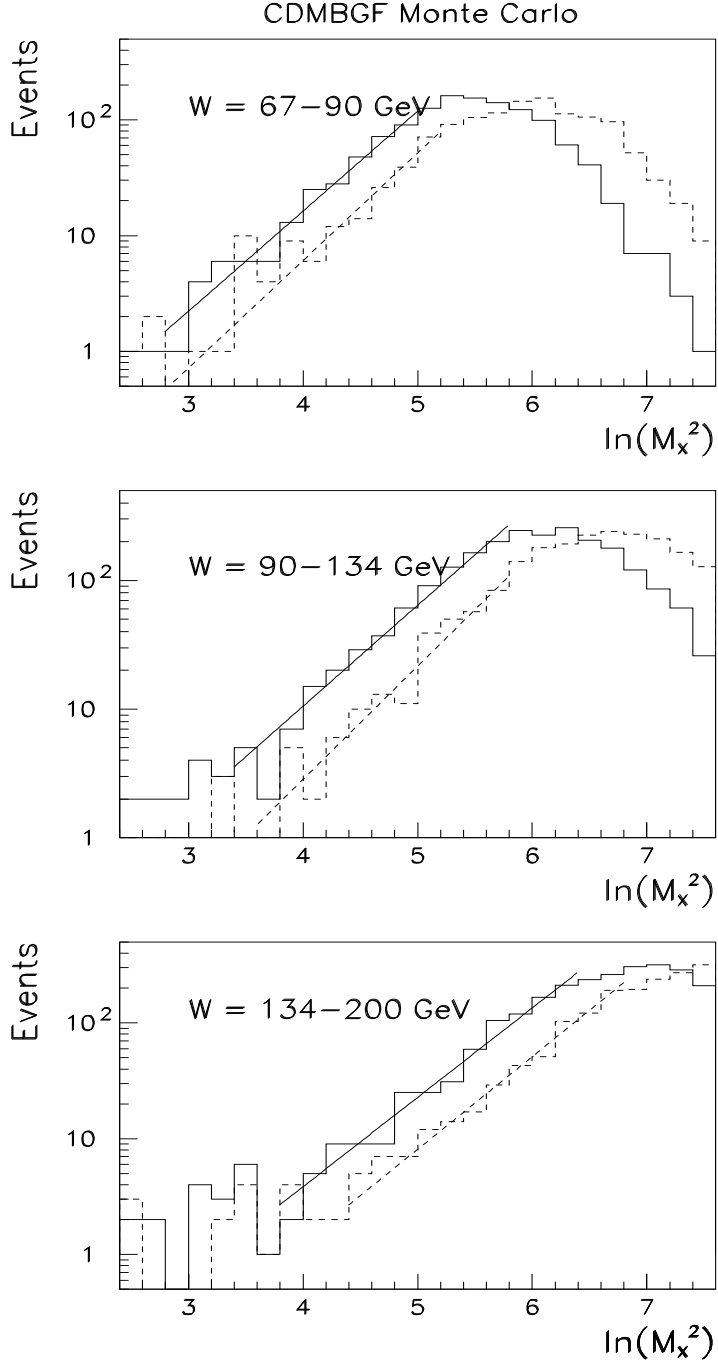}
\unitlength1cm
\begin{picture}(15,20)
\thicklines
\end{picture}
\caption{\label{f:cdmlogmadist}
        {    Comparison of $\ln M^2_X$ distributions for Monte Carlo events
             produced 
             with CDMBGF at the generator level (dashed histograms) and at 
             the detector level (full histograms) for $Q^2 = $14~GeV$^2$. 
             At the generator level,
             $M_X$ denotes the true mass of the particle system produced up
             to pseudorapidities of $\eta = $4.3 (forward edge of calorimeter). 
             At the detector level $M_X$ denotes the observed and
             uncorrected mass. The straight dashed and
             full lines show the fits of the exponential slopes to the
             CDMBGF distributions.}}
\end{figure}

We would like to add three remarks. Firstly, the value of the slope $b$ is
little affected by detector effects: it is almost the same at the detector
level as at the generator level. This was verified by Monte Carlo simulation
of nondiffractive events with CDMBGF. The MC events were selected using the
same selection cuts as for the data. The mass $M_X$ of a standard
nondiffractive DIS event at the generator level was defined as the invariant
mass of all particles (excluding the scattered electron) generated with
pseudorapidities $\eta <4.3$, the nominal end of the detector. In
Fig.~\ref{f:cdmlogmadist} the MC generated (dashed histograms) and MC
measured mass distributions (solid histograms) are shown. The exponential
slope values of $b = 1.9 \pm 0.1$ (shown as straight lines) obtained at the
generator level agree with the slope values found at the detector level to
within $\pm 0.1$ units compatible with the statistical errors of the
simulation. 

Secondly, the value predicted by CDMBGF for the slope is $b = 1.9 \pm 0.1$
while the data yield a shallower slope, the average being $\overline{b} =
1.46 \pm 0.15$. Note, the value of the slope $b$ cannot be predicted
precisely by the models; rather, DIS data have to be used to fix the
relevant parameters of the CDMBGF model.  For instance, short range
correlations arising from resonance production affect the value of $b$. The
observed difference between the exponential fall-off found in the data and
predicted by the Monte Carlo simulation indicates that estimating the tail
of the nondiffractive background from Monte Carlo simulation alone may lead
to an incorrect result for the diffractive cross section. 

Thirdly, we determine the slope $b$ from the data in a region where the
nondiffractive contribution dominates. The exponential fall-off will be
assumed to continue with the same slope into the region of overlap with the
diffractive contribution. The nondiffractive event sample generated with
CDMBGF indicates a small excess of events above the exponential fall-off
(see above). If allowance is made for a similar deviation from the
exponential fall-off in the data, the numbers of diffractive events obtained
after subtraction of the nondiffractive background change by less than
$\approx 30\%$ of their statistical error. 

\subsection{$M_X$ dependence of the diffractive contribution}

In diffractive events, the system $X$ resulting from the dissociation of the
photon is, in general, almost fully contained in the detector while the
outgoing proton or low mass nucleonic system, $N^{dissoc}$, escapes through
the forward beam hole. Furthermore, diffractive dissociation prefers small
$M_X$ values and leads to an event distribution of the form $d{\cal
N}/dM_X^2 \propto 1/(M_X^2)^{(1+n)}$ or
\begin{eqnarray} 
\frac{d{\cal N}}{d \ln M_X^2} \sim \frac{1}{(M_X^2)^n},
\label{dif}
\end{eqnarray}
approximately independent of $W$. At high energies and for large $M_X$ one
expects $n \approx 0$ leading to a roughly constant distribution in $\ln
M^2_X$. Such a mass dependence is seen in diffractive dissociation of $pp$
scattering (see e.g.~\cite{Goulian1,Cdf94}). A value of $n \approx 0$ is
also expected in diffractive models as the limiting value for the fall-off
of the mass distribution (see e.g. the NZ model~\cite{Nz}). 

To summarize this and the previous section, the diffractive contribution is
identified as the excess of events at small $M_X$ above the exponential
fall-off of the nondiffractive contribution with $\ln M^2_X$. The
exponential fall-off permits the subtraction of the nondiffractive
contribution and therefore the extraction of the diffractive contribution
without assuming the precise $M_X$ dependence of the latter. The $M_X$
distribution is expected to be of the form
\begin{eqnarray} 
\frac{d{\cal N}}{d\,\ln M_X^2} = D +  c\, \exp (b\,\ln {M_X^2}), ~~~~
{\rm for} ~~~ \ln M_X^2 \leq \ln W^2 -\eta_0. 
\label{eq:diffnondiff}
\end{eqnarray}
Here, $D$ denotes the diffractive contribution, the second term represents
the nondiffractive contribution and $\ln W^2 -\eta_0$ is the maximum value
of $\ln M^2_X$ up to which the exponential behaviour of the nondiffractive
part holds. We shall apply Eq.~\ref{eq:diffnondiff} in a limited range of
$\ln M^2_X$ for fitting the parameters $b, c$ of the nondiffractive
contribution. The diffractive contribution will not be taken from the fit
result for $D$ but will be determined by subtracting from the observed
number of events the nondiffractive contribution found in the fits. 
 
\subsection{Contribution from nucleon dissociation}
The contribution from the diffractive process where the nucleon dissociates,
\begin{displaymath} 
ep \to e + X  + N^{dissoc},
\end{displaymath}
was estimated by Monte Carlo simulation.  Assuming factorisation and a
Triple Regge formalism ~\cite{Mueller,Fiefox} for modelling $\gamma^* p \to
X + p$ and $\gamma^* p \to X + N^{dissoc}$, the measured cross sections for
elastic and single diffractive dissociation in $pp$ (and $\overline{p}p$)
scattering, $pp \to p + p$ and $pp \to p + N^{dissoc}$, were used to relate
$\gamma^* p \to X + N^{dissoc}$ to $\gamma^* p \to X + p$. 

The secondary particles from $N^{dissoc}$ decay were found to be strongly
collimated around the direction of the proton beam. Analysis of the angular
distribution of the secondary particles as a function of $M_N$ showed that
for $M_N<2$~GeV basically no energy is deposited in the calorimeter while
for events with $M_N>6$~GeV there are almost always secondaries which
deposit energy in the calorimeter. Furthermore, events of the type $ep \to e
+ X+ N^{dissoc}$, where decay particles from $N^{dissoc}$ deposit energy in
the calorimeter, have in general a mass reconstructed from all energy
deposits in the calorimeter (including those from the decay of the system
$X$ but excluding those of the scattered electron) which is much larger than
the mass of $X$. As a result these events make only a small contribution to
the event sample selected below for diffractive production of $\gamma^* p
\to X + N$ either with $M_X < 7.5$~GeV or $M_X < 15$~GeV. To a good
approximation, the selection includes all events from dissociation of the
nucleon, $\gamma^*p \to X + N^{dissoc}$, with $M_N < 4$~GeV. From the
comparison with the $pp$ data, we estimate the contribution of $\gamma^*p
\to X + N^{dissoc}$ with $M_N < 4$~GeV to the diffractive sample to be $11
\pm 5 \%$. 
  
\section{Extraction of the diffractive contribution}  

\subsection{Binning in $Q^2$ and $W$}
\label{s:binning}
The cross section for $\gamma^* p \to X + N$ was determined for two $Q^2$
intervals, $10$ - $20$~GeV$^2$ and $20$ - $56$~GeV$^2$, the average values
being 14~GeV$^2$ and 31~GeV$^2$, respectively. This choice of $Q^2$
intervals was motivated by the available event statistics and by the
requirement of good acceptance. These requirements also determined the $W$
range of $60 < W < 245$~GeV.  The intervals in $W$ were chosen so as to have
equidistant bins in $\ln W^2$ providing approximately equal numbers of
events in each $W$ bin.  For the bin width, $\Delta$ $\ln W^2$ = 0.4 was
used, commensurate with the resolution for $\ln W^2$ of 0.17 for diffractive
events and 0.32 for nondiffractive events, as determined by Monte Carlo
simulation.  The better resolution for diffractive events is due to the fact
that here the decay particles from the system $X$ are almost completely
contained in the detector. 

The total number of accepted events in the region $60 < W < 245$~GeV
and $10<Q^2<20\,{\rm GeV}^2$ ($20<Q^2<56\,{\rm GeV}^2$) was 14466 (11247).

\subsection{Fitting the nondiffractive contribution}

\begin{figure}[ht]
\includegraphics{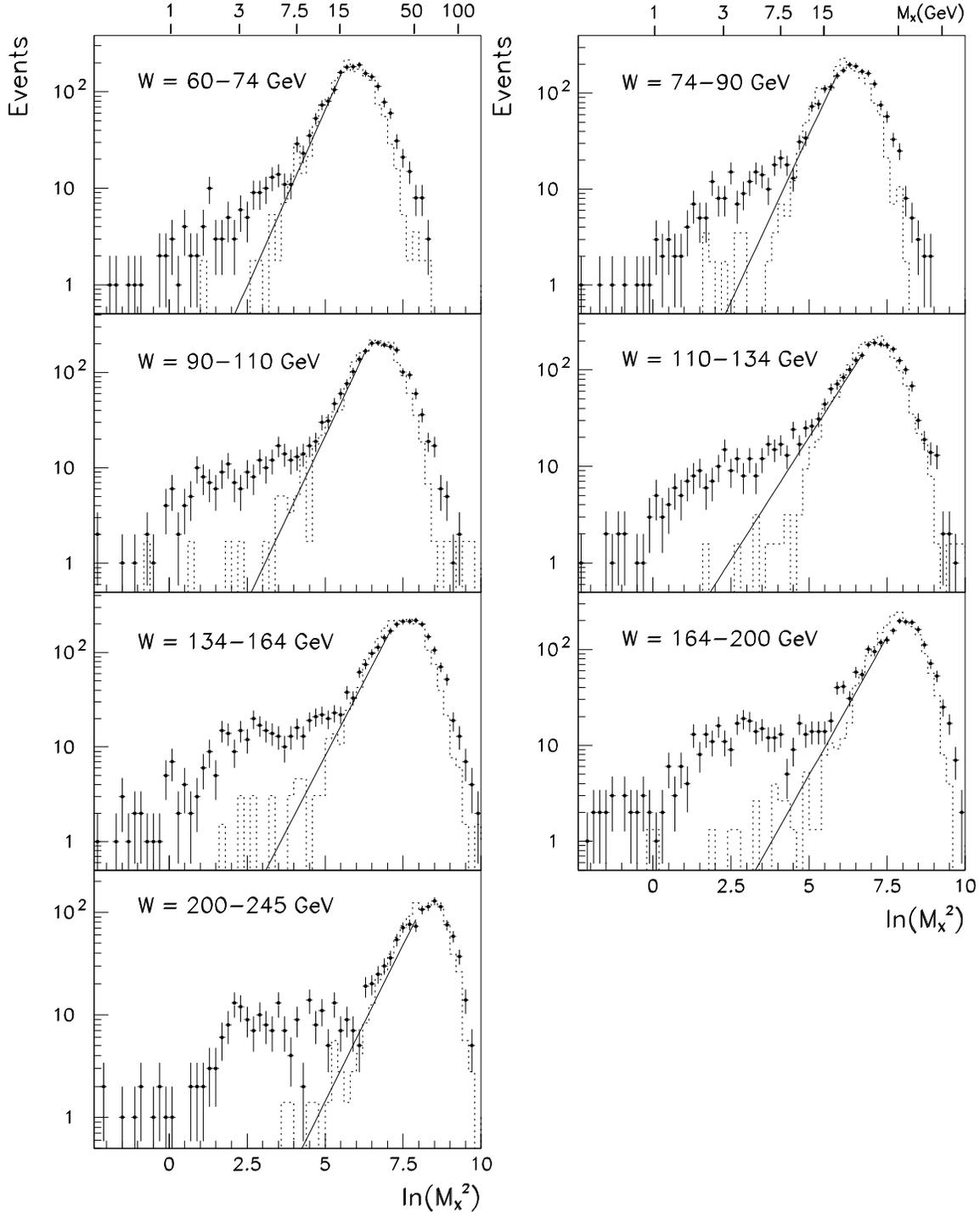}
\unitlength1cm
\begin{picture}(15,20)
\thicklines
\end{picture}
\caption{\label{f:logmadista}
        {    Distributions of $\ln M^2_X$ for the $W$  
             intervals indicated at $Q^2 = 14$~GeV$^2$. The data are shown
             with error bars which give the statistical errors.  
             Here $M_X$ is the corrected mass. The distributions are 
             not corrected for detector effects. The solid lines
             show the extrapolation of the nondiffractive background 
             as determined by the fits (see text). The dotted histograms
             show the predictions for nondiffractive scattering as
             calculated from CDMBGF. The CDMBGF distributions were normalized
             to 85$\%$ of the number of events in the data.}}
\end{figure}

\begin{figure}[ht]
\includegraphics{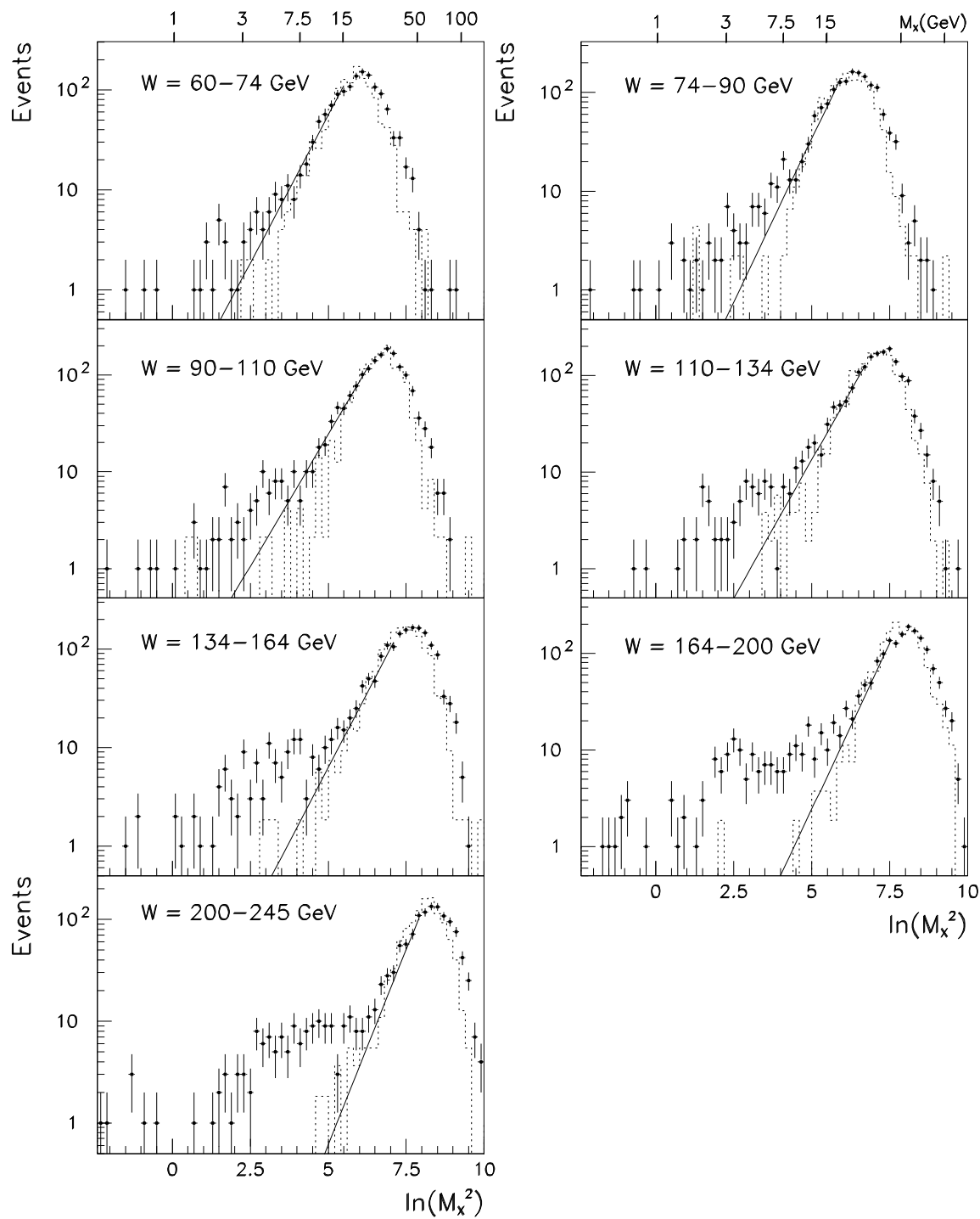}
\unitlength1cm
\begin{picture}(15,20)
\thicklines
\end{picture}
\caption{\label{f:logmadistb}
        { Distributions of $\ln M^2_X$ for the $W$ intervals indicated
          at $Q^2 = 31$~GeV$^2$. See caption of previous figure.}}
\end{figure}

\begin{figure}[ht]
\includegraphics{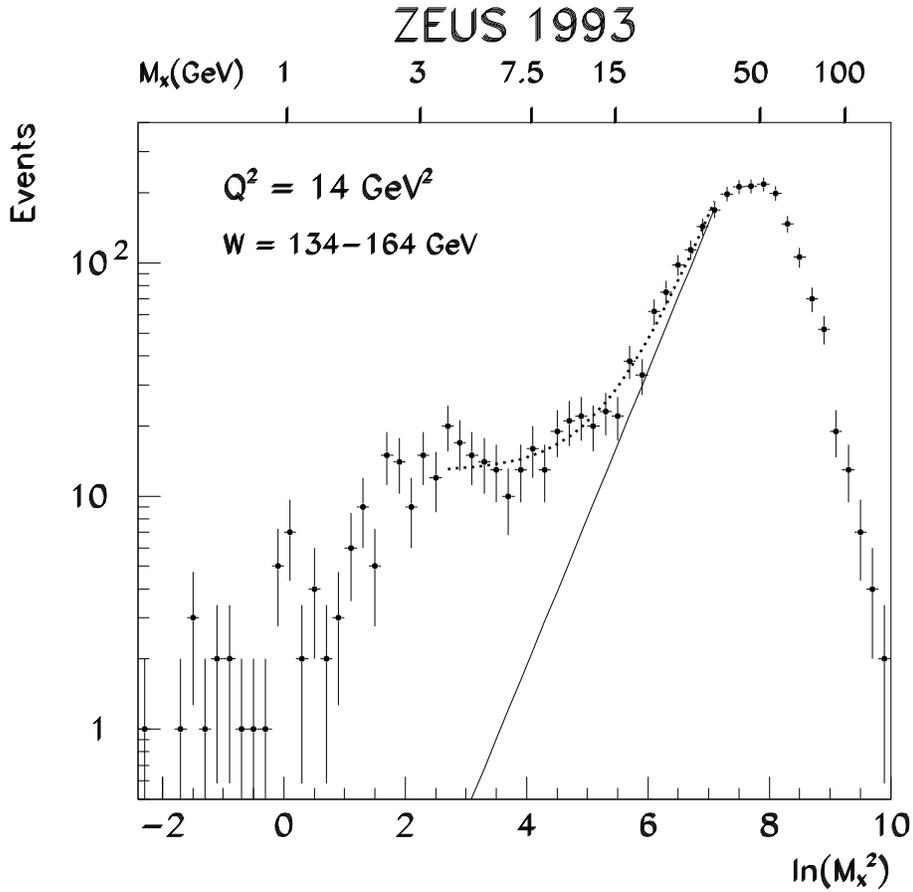}
\unitlength1cm
\begin{picture}(15,20)
\thicklines
\end{picture}
\caption{\label{f:fitexample}
        { 
             Example of a fit for the determination of the nondiffractive
             background in the $W$ interval 134-164~GeV at $Q^2 = $14~GeV$^2$.
             The data distribution of $\ln M^2_X$ is shown (uncorrected for
             detector effects) with error bars which give the 
             statistical errors. Here $M_X$ is the corrected mass.
             The dotted line shows the fit performed with $D_2 =$~0. 
             The beginning and the end of this 
             line show the $\ln M^2_X$ range over which the fit was performed. 
             The solid line shows the nondiffractive 
             background as determined by the fit (see text).}}
\end{figure}

The mass distributions for all $W$ and $Q^2$ intervals are presented in
Figs.~\ref{f:logmadista}~-~\ref{f:logmadistb} in terms of $\ln M_X^2$. The
diffractive contribution was obtained for each $W, Q^2$ interval by
determining first the nondiffractive background from a fit to the data at
large $\ln M^2_X$. The nondiffractive background was then extrapolated to
the small $\ln M^2_X$ region and subtracted from the data giving the
diffractive contribution as illustrated in Fig.~\ref{f:fitexample}. The fit
for the nondiffractive background was performed using
Eq.~\ref{eq:diffnondiff}. The diffractive contribution $D$ was assumed to be
of the form 
\begin{eqnarray}
 \frac{d{\cal N}^{diff}}{d\,\ln M_X^2} = D_1 \;  + \;  \frac{D_2}{M^2_X}.
\label{eq:diffmodif}
\end{eqnarray}
where $D_1$ and $D_2$ are constants. The second term allows for
contributions from lower lying Regge poles contributing to $\gamma^*$
pomeron scattering at a c.m. energy of $M_X$ or from a hard quark
distribution in the pomeron (e.g. Hard POMPYT). The fit parameters are
$D_1$, $D_2$, $b$ and $c$. 

The fits were performed to the data in the range $\ln Q^2 < \ln M^2_X < {\rm
Max} (\ln M^2_X)$. The lower limit for $\ln M^2_X$ was chosen according to
the expectation of the diffraction models ~\cite{pompyt,Nz,Donlan2}, that
for $M^2_X > Q^2$ the diffractive contribution is of the form given by
Eq.~\ref{eq:diffmodif}. The upper limit Max$(\ln M^2_X)$ was chosen as the
maximum value of $\ln M^2_X$ up to which the data exhibit an exponential
behaviour. The maximum value of $\ln M^2_X$ was determined by fitting the
$\ln M^2_X$ distributions for each ($W, Q^2$) interval with a varying
maximum value of $\ln M^2_X$. In most ($W, Q^2$) intervals a clear boundary
as a function of Max$(\ln M^2_X)$ was observed beyond which the $\chi^2$
probability for the fit dropped rapidly. The boundary marks the location
where the distribution starts to deviate from an exponential behaviour. The
boundary was found to be at $\ln M^2_X + \ln (s/W^2) = 8.8$ corresponding to
a value of $\eta_0 = 3.0$, a value which is in good agreement with
expectations of $\eta_0$~= ${\cal Y}_{max} - {\cal Y}^{det}_{limit}$ (see
discussion above).  The fits were performed with Max$(\ln M^2_X) = \ln
(W^2/s) + 8.6$. In the studies of systematic uncertainties the maximum value
was increased (decreased) by 0.2 (0.4), see section~\ref{s:systematics}. The
$\chi^2$ probabilities for the fits were on average 40$\%$.
 
The fits were performed by including both terms, $D_1$ and $D_2$, (extended
fits) as well as by assuming $D_2 = 0$ (nominal fits). Since the two fit
procedures resulted in only minor differences we used the results from the
nominal fits and used those from the extended fits for estimating the
systematic errors (see section~\ref{s:systematics}).  The solid lines in
Figs.~\ref{f:logmadista}~-~\ref{f:logmadistb} show for all $Q^2$ and $W$
bins the exponential fall-off of the nondiffractive contribution resulting
from the fits. The nondiffractive contribution moves to larger $\ln M^2_X$
values proportional to $\ln W^2$. As $W$ increases, the diffractive
contribution appears with little background over an increasing $M_X$ range.
Above $W = 90~(164)$~GeV the nondiffractive background is small up to $M_X$
values of 7.5 (15)~GeV (see also Table~1 below). 

 In Figs.~\ref{f:logmadista} and \ref{f:logmadistb} the nondiffractive
background estimates are also compared with the predictions of the CDMBGF
simulation. The dotted histograms display the predictions of CDMBGF for the
nondiffractive contribution normalized to 85$\%$ of the number of events
observed in the data in each ($W, Q^2$) bin, the 15\% reduction accounting
roughly for the diffractive contribution. The qualitative features of the
data are reproduced by CDMBGF in all $Q^2$ and $W$ intervals, although the
exponential slopes are steeper than in the data (see discussion in
section~\ref{s:lpsnondiff}). 

\section{Determination of the diffractive cross section}
\label{s:detercross}
The number of diffractive events, ${\cal N}^{diff}_{meas}$, was determined
in all $Q^2$ and $W$ bins for the $M_X$ intervals $ < 3$~GeV ($\ln M_X^2 <
2.2$), $3 - 7.5$~GeV ($\ln M_X^2 = 2.2 - 4.0$) and $7.5 - 15$~GeV ($\ln
M_X^2 = 4.0 - 5.4$) by subtracting from the observed number of events,
${\cal N}_{obs}$, the contribution from electron beam gas scattering, ${\cal
N}^{e gas}$, and the nondiffractive contribution, ${\cal N}^{nondiff}$,
obtained from the fit, ${\cal N}^{diff}_{meas} = {\cal N}_{obs}-{\cal N}^{e
gas}-{\cal N}^{nondiff}$.  Electron-gas scattering produces low $M_X$
events, which are candidates for diffractive scattering;  most of them have
no reconstructed vertex. Events from unpaired electron bunches were used to
obtain ${\cal N}^{e gas}$. 

From ${\cal N}^{diff}_{meas}$ the number of produced diffractive events,
${\cal N}^{diff}_{prod}$, was obtained by a Bayesian unfolding procedure
which took into account detector effects such as bin-to-bin migration,
trigger biases and event selection cuts. In the unfolding, the event
distribution generated by Monte Carlo, $n_G(i)$, was reweighted as discussed
in section~\ref{s:Weighting} so that the observed Monte Carlo distribution,
$n_O(j)$, reproduced closely ${\cal N}^{diff}_{meas}$, the event
distribution measured in the data which will be denoted by $n_{Dat}(j)$. The
indices $i, j$ denote the three - dimensional bins in ($M_X$, $W$, $Q^2$) in
which the data were analyzed.  For every set of weights we determined from
the generated distribution the observed distribution and the transfer matrix
$T_{GO}(j,i)$, which leads from the observed to the generated distribution,
$ n_G(i) = \sum_j T_{GO}(j,i) n_O(j)$. In an iterative procedure the
$\chi^2$ obtained from the differences of $n_{Dat}(j)$ and $n_O(j)$ was
minimized by varying the weighting parameters of the TRM function given in
Eq.~\ref{eq:TRMweight}. A good description of the data was obtained; the
minimum $\chi^2 = 34$ for $28$ degrees of freedom. The values found for the
weighting parameters were $c_L = 0.1$, $M_0/Q = 1$ and $\overline{\alphapom}
= 1.2$. The resulting matrix $T_{GO}(j,i)$ was then used to determine the
unfolded event distribution from that measured, $n_U(i) = \sum_j T_{GO}(j,i)
n_{Dat}(j)$. The number of unfolded events is denoted by ${\cal
N}^{diff}_{prod}$. In the calculation of the statistical errors, the
bin-to-bin correlations were neglected. The errors were checked by examining
the spread of results obtained by dividing the data into several subsamples.

\subsection{Evaluation of the cross sections} For the final analysis, only
bins where the fraction of nondiffractive background was less than 50~$\%$
and the purity was above 30~$\%$ were kept. Purity is defined as the ratio
of the number of events generated in the bin and observed in the same bin
divided by the total number of events observed in the bin. The average
purity in a ($M_X, W, Q^2$) bin was 43~$\%$. 

Electromagnetic radiative effects were corrected for in every ($M_X, W,
Q^2$) bin~\cite{Wulff}. The corrections were less than $10\%$ and
independent of $W$. 

The average differential cross section for $ep$ scattering, in a ($M_X, W,
Q^2$) bin, is obtained by dividing the number of unfolded events, ${\cal
N}^{diff}_{prod}$, by the luminosity and the bin widths. The lower limit of
$M_X$, was taken to be 2$m_{\pi}$, where $m_{\pi}$ is the pion mass.

The cross sections for the process $ep \to e X N$ can be expressed in terms of 
the transverse (T) and longitudinal (L) cross sections $\sigma_T^{diff}$,
$\sigma_L^{diff}$, for $\gamma^* p \to X N$ as follows~\cite{Hand}:
\begin{eqnarray}
 Q^2\,\frac{d\sigma^{diff}_{ep \to eXN}(M_X,W,Q^2)}{dM_X d\ln W^2 dQ^2} =
 \frac{\alpha}{2\pi}\, [(1-y)^2 +1][\sigma_T^{diff}+\sigma_L^{diff}]
[1-\frac{y^2}{(1-y)^2+1}
\frac{\sigma_L^{diff}}{\sigma_T^{diff}+\sigma_L^{diff}}].  \label{stsl}
\end{eqnarray} The relative contribution of the correction term in the third
square bracket on the r.h.s. of Eq.~\ref{stsl} is negligible if $y \ll 1$ or
$\sigma^{diff}_L \ll \sigma^{diff}_T$. Since $y\approx W^2/s$, the
contribution can be substantial only at high $W$ values. In the extreme case
that $\sigma^{diff}_L \gg \sigma^{diff}_T$, the correction term will
increase $[\sigma^{diff}_T + \sigma^{diff}_L ]$ by at most $35\%$ for the
highest $W$ bin ($200 - 245$~GeV) and $11\%$ for the next highest $W$ bin
(164 - 200 GeV).  If $\sigma^{diff}_L = Q^2/M^2_X \cdot \sigma^{diff}_T$, as
e.g. in the Vector Dominance Model or in partonic models (see
e.g.~\cite{Frankfurtetal}), then the second term increases $[\sigma^{diff}_T
+ \sigma^{diff}_L ]$ by at most $31\%$ ($17\%, 6\%$) for the bins with the
highest $W$, $Q^2$ values and $M_X < 3$~GeV ($3 - 7.5$~GeV, $7.5 - 15$~GeV)
falling below $10\%$ ($6\%, 2\%$) in the next highest $W$ bin
\footnote{Diffractive $\rho^o$ production via $ep \to e \rho^o p$ is a known
contribution to $\sigma^{diff}_L$ (see ~\cite{Zeprho}). The fraction of
diffractive events in the lowest $M_X$ bin ($2m_{\pi} < M_X < 3$~GeV) from
$\rho^o$ production is around 20 $\%$ for the $W$, $Q^2$ region under
study.}. 

In the following analysis, the correction term was set equal to one:
\begin{eqnarray}
\frac{d\sigma^{diff}_{\gamma^*p \to XN}(M_X,W,Q^2)}{dM_X}  
\equiv \frac{d(\sigma_T^{diff}+\sigma_L^{diff})}{dM_X} \approx
\frac{2\pi}{\alpha}\,\frac{Q^2}{ (1-y)^2 +1}
 \,\frac{d\sigma^{diff}_{ep \to eXN}(M_X,W,Q^2)}{dM_X d\ln W^2 dQ^2} \; .
\label{sgp} 
\end{eqnarray}
The numbers of events observed and estimated to come from background are
given in Table~1, together with the values of the $ \gamma^*p \to XN$
differential cross sections averaged over the specified $M_X$ range. The
cross sections are quoted for the $Q^2$ values of 14 and 31 GeV$^2$ and the
$W$ values corresponding to the logarithmic means of the $W$ interval
limits. The average $M_X$ values in each $M_X$ interval are 1.9, 5.1, 11.0
GeV at $Q^2=14$ GeV$^2$ and 2.0, 5.1 and 11.0 GeV at $Q^2=31$ GeV$^2$.

\subsection{Systematic errors}
\label{s:systematics}

The systematic errors on the cross sections were estimated by varying the
cuts and algorithms used to select the events. The bin-by-bin changes from
the standard values were recorded. For reference each test is numbered; the
number is given in brackets \{\}. 

The efficiency for finding the scattered electron was around $100\%$ for
$E^\prime_e > 18$~GeV, falling to $55\%$ at $E^\prime_e = 10$ GeV.  To
evaluate the resulting uncertainty on the cross section the cut on
$E^\prime_e$ was raised~\{1\} (lowered~\{2\}) from 8 to 10~GeV (7~GeV); the
box cut was varied from 32~cm to 36~cm~\{3\} (28~cm~\{4\}). The changes on
the measured cross sections were negligible except for one low $W$ bin where
a $28\%$ change of the cross section was observed. 

In order to test for remaining background from photoproduction and for the
sensitivity to radiative effects, which were not simulated in the
diffractive Monte Carlo, the cut on $\sum_h (E - p_Z)$ was raised~\{5\}
(lowered~\{6\}) from $35$~GeV to $38$~GeV ($32$~GeV). This resulted in small
changes of the cross sections, except in one bin where it reached $20\%$. 

The effect of the cut on $y_{JB}$ was tested by raising~\{7\}
(lowering~\{8\}) it from $0.02$ to $0.03$ ($0.01$). The changes were found
to be below $5\%$, except for the low $M_X$ interval where in two low $W$
bins the higher $y_{JB}$ cut substantially reduced the acceptance; for these
bins changes of $19\%$ and $47\%$, respectively, were observed. 

The mass correction factor was assumed to be different by +10$\%$~\{9\}
(-10$\%$~\{10\}) in the Monte Carlo simulation and in the data. This
affected the low $W$ bins where changes up to $17\%$ were observed. 

The analysis was performed without requiring a reconstructed event vertex
and could therefore be affected by background from beam-gas scattering.  The
analysis was repeated requiring an event vertex as determined by tracking
where the Z-coordinate of the vertex had to lie in the interval -50~cm to
+40~cm~\{11\}. This requirement was satisfied by $89\%$ of the accepted
events. The vertex requirement reduced the total number of events in the
accepted ($M_X$,$W$,$Q^2$) bins obtained with unpaired electron bunches from
12 to 1 event thereby suppressing the electron-gas background to a
negligible amount.  Apart from this effect reductions of the cross sections
by at most $12(16)\%$ in the highest $W$ bins at $Q^2 = 14(31)~$GeV$^2$ were
observed. For the other bins the changes were smaller. 

Uncertainties in the number of diffractive events resulting from the
subtraction of the nondiffractive background were estimated by
increasing~\{12\} (decreasing~\{13\}) the upper limit of the fit range in
$\ln M^2_X$ by 0.2 (0.4) units; by decreasing the lower limit of the fit
range in $\ln M^2_X$ by 0.4 units~\{14\}; by repeating the fits for the
nondiffractive background with the extended form Eq.~\ref{eq:diffmodif},
$D_2~\neq~0$, for the diffractive part~\{15\}.  The typical changes were
below 10$\%$. The largest change was observed for the lowest $W$ bin at
$Q^2=31$~GeV$^2$ where it reached $37\%$. 

The uncertainty resulting from the Monte Carlo weighting procedure was
estimated by varying the weighting parameter $\overline{\alphapom}$ from 1.2
to 1.1~\{16\} and 1.3~\{17\}. To estimate the uncertainty of the Monte Carlo
modelling for the low $M_X$ region the $\rho^o$ events for $M_X < 1$ GeV
were replaced by a system decaying into $\pi^+$, $\pi^-$, $\pi^o$'s with a
phase space like distribution.  Changes of less than $8\%$ were observed. 

The total systematic error for each bin was determined by adding
quadratically the individual systematic uncertainties, separately for the
positive and negative contributions. The total errors were obtained by
adding the statistical and systematic errors in quadrature.  The errors do
not include an overall normalization uncertainty of 3.5$\%$ of which 3.3$\%$
is from the luminosity determination and 1$\%$ from the uncertainty in the
trigger efficiency.

\section{Differential cross section for $\gamma^*p \to X + N$}

\begin{figure}[ht]
\includegraphics{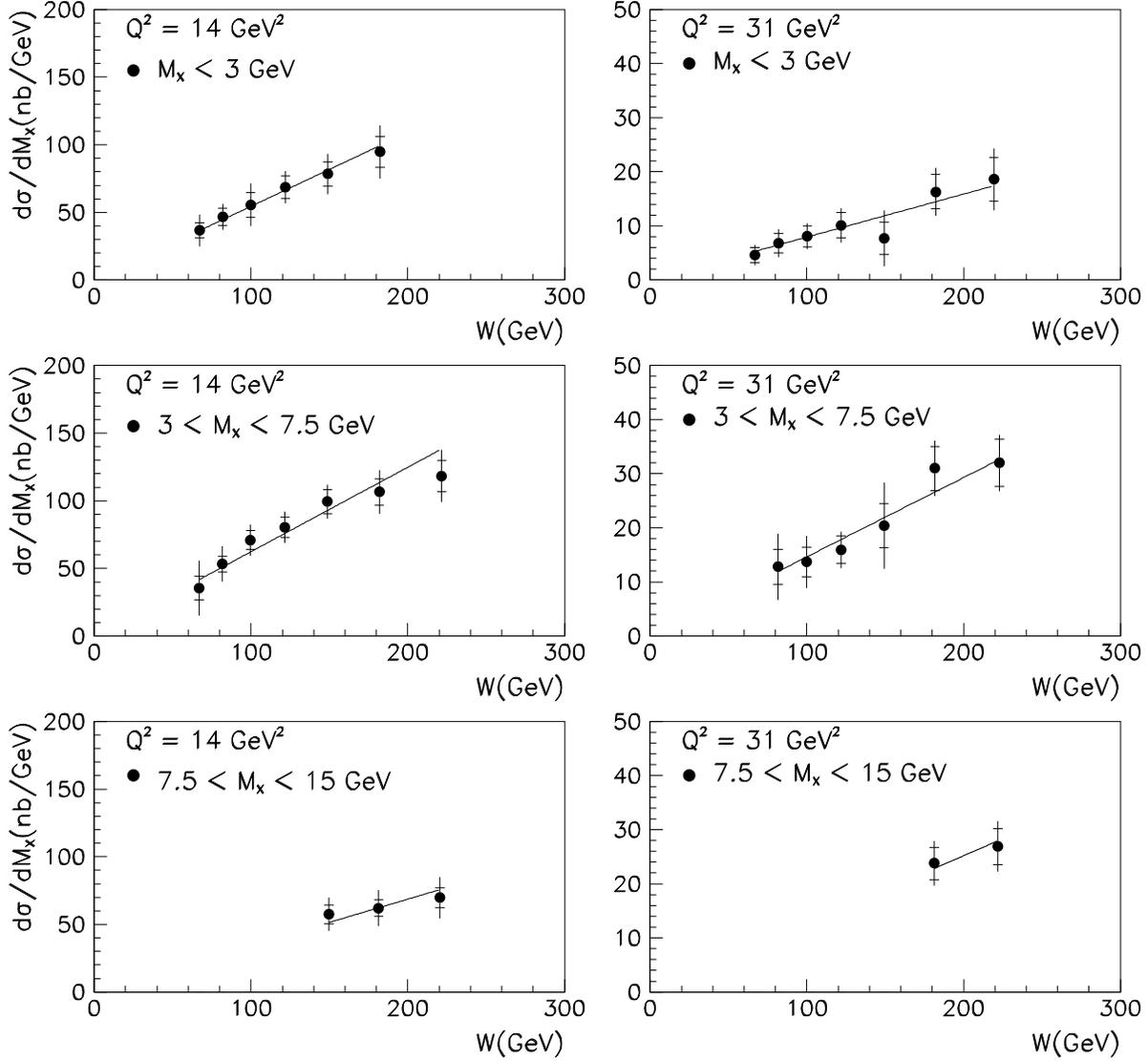}
\unitlength1cm
\begin{picture}(15,20)
\thicklines
\end{picture}
\caption{\label{f:dsigdmxvsw}
        {    The differential cross sections $d\sigma^{diff}
             (\gamma^* p \to X N)/dM_X$ as a function of $W$ averaged over
             the $M_X$ intervals 2$m_{\pi}$ - 3, 3 - 7.5 and 7.5 - 15~GeV 
             at $Q^2$ = 14 and 31~GeV$^2$. The inner error bars show the 
             statistical errors and the full bars the statistical and
             systematic errors added in quadrature.  The overall normalization 
             uncertainty of 3.5$\%$ is not included. 
             The curves show the results 
             from fitting all cross sections to the form  $d\sigma^{diff}/dM_X 
             \propto (W^2)^{(2\overline{\alphapom}-2)}$ with a common value of 
             $\overline{\alphapom}$, see text.}}
\end{figure}

The differential cross section $d\sigma^{diff}_{\gamma^* p \to XN}/dM_X$ was
determined according to Eq.(\ref{sgp}) for the different ($M_X$,$W$,$Q^2$)
intervals. The results are shown in Fig.~\ref{f:dsigdmxvsw}, as a function
of $W$ averaged over the $M_X$ bins $2 m_\pi $ - 3~GeV, 3 - 7.5~GeV and 7.5
- 15~GeV at $Q^2 = 14$ and $31$~GeV$^2$. The solid points show the measured
values. The inner error bars give the statistical errors. For the full bars
the statistical and systematic errors have been added in quadrature. 

The cross section, within errors, is seen to rise linearly with $W$ at both
$Q^2$ values for all $M_X$ bins up to 7.5~GeV. For the $M_X$ bins (7.5 -
15)~GeV the data in the accepted $W$ range are consistent with this rise. 

\begin{figure}[ht]
\includegraphics{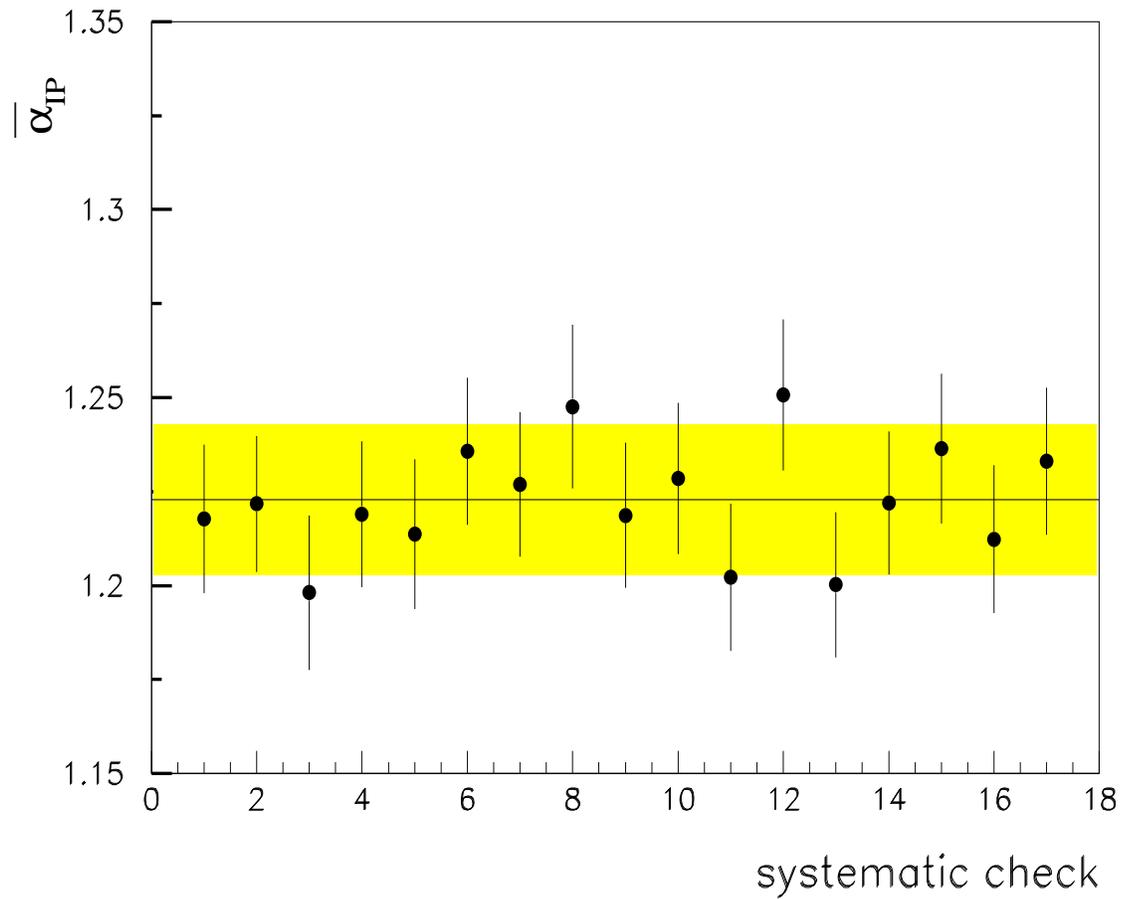}
\unitlength1cm
\begin{picture}(15,20)
\thicklines
\end{picture}
\caption{\label{f:alpsys}
        {    Sensitivity of the value of $\overline{\alphapom}$ to the 
             different sources of systematic uncertainties.
             The central line and the shaded band show for the standard fit
             the value of $\overline{\alphapom}$ and $\pm$~1~s.d.
             The dots give the $\overline{\alphapom}$ value with its 
             uncertainty obtained by repeating the analysis for each 
             systematic check labeled 1 through 17 as described in 
             the text.}}
\end{figure}

In a Regge - type description ~\cite{Mueller,Fiefox},
the $W$ dependence of the diffractive cross section is of the form
\begin{eqnarray}
\frac {d\sigma^{diff}_{\gamma^*p \to XN}(M_X,W,Q^2,t)}{ dt dM_X } & \propto
(W^2)^{2\alphapom(0) -2} \; \cdot \;  e^{t(B+2\alphapom^\prime\ln
(W^2/(M_X^2+Q^2))}\; \; \; ,
\label{eq:gxp4}
\end{eqnarray}
where $\alphapom(t)$ = $\alphapom(0)$ + $\alphapom^\prime t$ is the pomeron
trajectory and $B$ and $\alphapom^\prime $ are parameters. The cross
sections in each ($M_X$, $Q^2$) interval were fitted to the form
\begin{eqnarray}
\frac {d\sigma^{diff}_{\gamma^*p \to XN}(M_X,W,Q^2)}{ dM_X } & \propto
(W^2)^{(2\overline{\alphapom} -2)} \; \; \; ,
\label{eq:gxpint}
\end{eqnarray}
where $\overline{\alphapom}$ stands for $\alphapom(t)$ averaged over the $t$
distribution. The fit was performed by considering $\overline{\alphapom}$
and the six normalization constants for the six ($M_X$, $Q^2$) intervals as
free parameters.  Taking into account only the statistical errors,
$\overline{\alphapom}$ was found to be $1.23 \pm 0.02$ with $\chi^2 /ndf =
11.7/24$. The systematic uncertainties were estimated by repeating the fit
independently for every source of systematic error discussed above. The
results are shown in Fig.~\ref{f:alpsys}.  The lowest $\overline{\alphapom}$
value obtained was 1.20, the highest value was 1.25. The observed deviations
were added in quadrature leading to the final value: 

$$  \overline{\alphapom}\; \; = \; \; 1.23 \pm 0.02 (stat) \pm 0.04 (syst). $$

The $W$ dependence was also determined by restricting the analysis to those
($M_X$,$W$,$Q^2$) bins where almost all events have a large rapidity gap and
where, therefore, nondiffractive contributions should be negligible.  Of the
events observed with $M_X <$~3~GeV, 98$\%$ have a rapidity gap $\Delta\eta
>$~2.  The fit to these events yielded
$\overline{\alphapom}=$~1.24~$\pm~0.03~^{+0.07}_{-0.03}$.  Note that the
average rapidity gap for these events grows with $W$ and a possible
background from nondiffractive scattering would diminish with rising $W$.
Consequently, if there were a nondiffractive contribution left in the
sample, the correction for this background would lead to an increase of the
value of $\overline{\alphapom}$ compared to the result obtained. 

The effect of the kinematically allowed minimum value of $\vert t \vert $
($\vert t \vert_{min}$) on the $t$ - integrated cross section and therefore
on the value of $\overline{\alphapom}$ is negligible, $ \vert t \vert_{min}$
being less than $10^{-3}$~GeV$^2$. If $\alphapom^\prime = 0$ then
$\overline{\alphapom}$ is equal to the pomeron intercept at $t = 0$,
$\alphapom(0) = \overline{\alphapom}$. If the slope $\alphapom^\prime$ in
diffractive DIS is the same as for the soft pomeron ($\alphapom^\prime =
0.25$ GeV$^{-2}$) and the parameter $B$ is equal to half the value observed
for elastic $pp$ scattering~\cite{Goulian1} ($B=4.5$ GeV$^{-2}$), then
$\alphapom(0)$ will increase from 1.23 to 1.26. Our result can be compared
with the soft pomeron trajectory~\cite{Donlan1}, $\alphapom(t) = 1.08 + 0.25
t$, as determined from hadron~- hadron scattering.  Assuming $B
=$~4.5~GeV$^{-2}$ and averaging over $t$, the soft pomeron predicts
$\overline{\alphapom} = 1.05$. In extracting the diffractive cross section,
the assumption $\sigma^{diff}_L = 0$ was made. Assuming instead
$\sigma^{diff}_L = (Q^2/M^2_X) \sigma^{diff}_T$ (see for
example~\cite{Frankfurtetal}) will increase $\overline{\alphapom}$ from 1.23
to 1.28. Hence, a positive slope of the pomeron trajectory and/or a finite
$\sigma^{diff}_L$ contribution will lead to a larger $\alphapom(0)$ value
and increase the difference with respect to the soft pomeron intercept. 

The observation of $\overline{\alphapom}$ being substantially larger in
diffractive DIS than expected for the soft pomeron is in line with the
expectation of perturbative QCD ~\cite{Lipatov} and shows that deep
inelastic diffractive scattering has a perturbative contribution. The
measured value of $\overline{\alphapom}$ is smaller than the value which
would follow from the BFKL formalism, $\alphapom(0) \approx 1.5$. It is in
broad agreement with the effective pomeron intercept expected in the
perturbative models of ~\cite{Ellisetal,Buch} where the Bjorken-$x$
dependence of the gluon momentum density $xg(x,Q^2)$ of the proton
determines the $W$ dependence of diffractive scattering. 

There are also models where the effective $\alphapom(0)$ is expected to be
smaller in diffractive hadron~- hadron or photon~- hadron scattering as
compared to deep inelastic diffractive scattering because in hadron~- hadron
or photoproduction processes, in addition to single, multiple pomeron
exchanges also contribute (see e.g.~\cite{Capellaetal}). In these models the
importance of multiple pomeron exchanges decreases quickly with growing
photon virtuality. 
 
\subsection{Diffractive contribution to total deep inelastic scattering}

The relative contribution of diffractive scattering to the total
virtual-photon proton cross section, $ \sigma^{tot}_{\gamma^*p}$, was
determined for the $W$ bin (164 - 200~GeV) where data on the diffractive
cross section are available up to $M_X=15$~GeV.  The data for $
\sigma^{tot}_{\gamma^*p}$ were taken from the analysis of the proton
structure function $F_2$~\cite{Zepf2}. The ratio $r^{diff} = \int dM_X
\sigma^{diff}_{\gamma^*p \to XN}(M_N < 4~{\rm GeV}) /
\sigma^{tot}_{\gamma^*p}$ is given in Table~2 integrated over different
$M_X$ bins at $Q^2 = 14$ and $31$~GeV$^2$. In the lowest $M_X$ bin the
relative contribution from diffractive scattering to the total DIS drops by
a factor of about three from $Q^2 = 14$ to $31$~GeV$^2$. With increasing
$M_X$ the relative contributions from diffractive scattering tend to become
equal for the two values of $Q^2$. The observed $Q^2$ behaviour does not
preclude a leading twist behaviour of the diffractive DIS cross section
(observed by~\cite{Zeplrg,Heplrg}): the measurements for the two different
$Q^2$ values correspond to different values of $x$, namely $x = Q^2/W^2 = 4
\cdot 10^{-4}$ and $9 \cdot 10^{-4}$, respectively. Furthermore, it is
conceivable that for fixed $x$ the $Q^2$ behaviour of the diffractive cross
section changes with $M_X$ and only its integral over $M_X$ is of leading
twist~\cite{Stodolsky}. 

\section{Diffractive structure function of the proton}

The DIS diffractive cross section, $ep \to e + X + N$, can be expressed in
terms of the diffractive structure function $F^{D(3)}_2(\beta,Q^2,\xpom)$ as
follows~\cite{Ingel1}: 
\begin{eqnarray}
  \frac {d\sigma^{diff}_{ep \to eXN}(\beta,Q^2,\xpom,M_N<4\,{\rm GeV})}{
d\beta dQ^2 d\xpom } & = &
  \frac {2 \pi \alpha^2}{\beta Q^4} [1 + (1-y)^2] F^{D(3)}_2 (\beta,Q^2,\xpom) 
 \label{eq:epfd3}
\end{eqnarray}

if the contribution from longitudinal photons is neglected.

\begin{figure}[ht]
\includegraphics{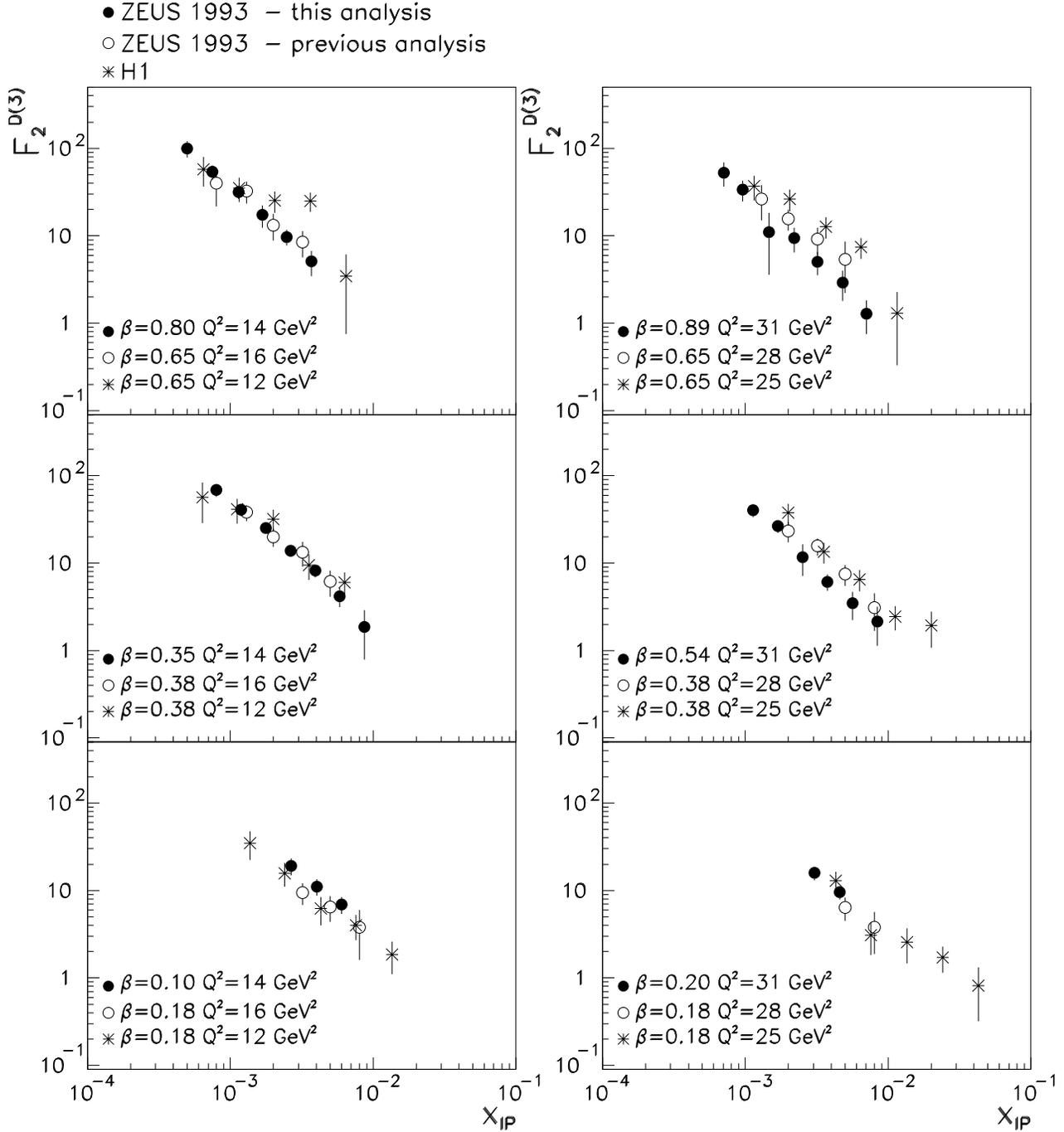}
\unitlength1cm
\begin{picture}(15,20)
\thicklines
\end{picture}
\caption{\label{f:fd3}
        {    The diffractive structure function $F^{D(3)}_2$ as a function 
             of $\xpom$  from this
             analysis (solid dots). The error bars show the statistical
             and systematic errors added in quadrature. Also shown are
             the results from our previous 
             measurement\protect\cite{Zepdifff} (open dots) and from H1
             \protect\cite{Hepdifff} (stars)
             obtained at slightly different $\beta$ and $Q^2$ values.}}
\end{figure}

In Fig.~\ref{f:fd3} we show $F^{D(3)}_2$ of this analysis as calculated from
the differential cross sections $d\sigma^{diff}_{\gamma^*p \to XN}/dM_X$,
$M_N <$~4~GeV. The result is plotted as a function of $\xpom$ for different
values of $\beta$ and $Q^2$ (solid points). The error bars show the
statistical and systematic errors added in quadrature. The data from our
previous analysis~\cite{Zepdifff} of $F^{D(3)}_2$ are also shown as the open
points. In the previous analysis, the diffractive contribution was
determined with a rapidity gap method using CDMBGF to estimate the
nondiffractive part.  Note that these data have been evaluated at somewhat
different $\beta$ and $Q^2$ values than in the present analysis. 

The diffractive structure function falls rapidly with increasing $\xpom$,
the $\xpom$ dependence being the same within errors in all $\beta$ and $Q^2$
intervals. A good fit to the data from this analysis (solid points) is
obtained with the form $F^{D(3)}_2 = constant \cdot (1/\xpom)^a$ yielding $a
= 1.46 \pm 0.04 (stat) \pm 0.08(syst)$.  Note, for fixed $Q^2$ and $\beta$
the $\xpom$ dependence of $F^{D(3)}_2$ is equivalent to the $W$ dependence
of $d\sigma^{diff}(\gamma^*p \to XN)/dM_X$ discussed above, the values of
$a$ and $\overline{\alphapom}$ being connected by the relation
$\overline{\alphapom} = (a + 1)/2$. 

The value of $a$ measured in the present analysis is somewhat higher than
the value of $1.30 \pm 0.08 (stat) ^{+ 0.08}_{-0.14} (syst)$ found in our
previous $F^{D(3)}_2$ analysis which can be understood by the way the
nondiffractive background was subtracted. To investigate the effect of the
different background estimates we subtracted the nondiffractive contribution
using CDMBGF as in the previous analysis and determined $a$ for the same
kinematic region \footnote{The previous analysis was limited to the region
$M_X > 2.8$ GeV.} in $\xpom$ and $\beta$. The result was $a = 1.28 \pm 0.04$
while the new method gave $a = 1.42 \pm 0.08$. In the $M_X > 3$ GeV region
the difference to the previous analysis is due to the new method of
estimating the nondiffractive background. For $M_X < 3$ GeV both methods in
this analysis gave the same result ($a = 1.48 \pm 0.06$) which is not
surprising since in this region the nondiffractive contribution is found to
be negligible in both methods. 

Figure~\ref{f:fd3} shows also the $F^{D(3)}_2$ values obtained by the H1
collaboration~\cite{Hepdifff}. H1 found $a = 1.19 \pm 0.06 \pm 0.07$, a
value which is smaller than the result obtained in this analysis. 

\begin{figure}[ht]
\includegraphics{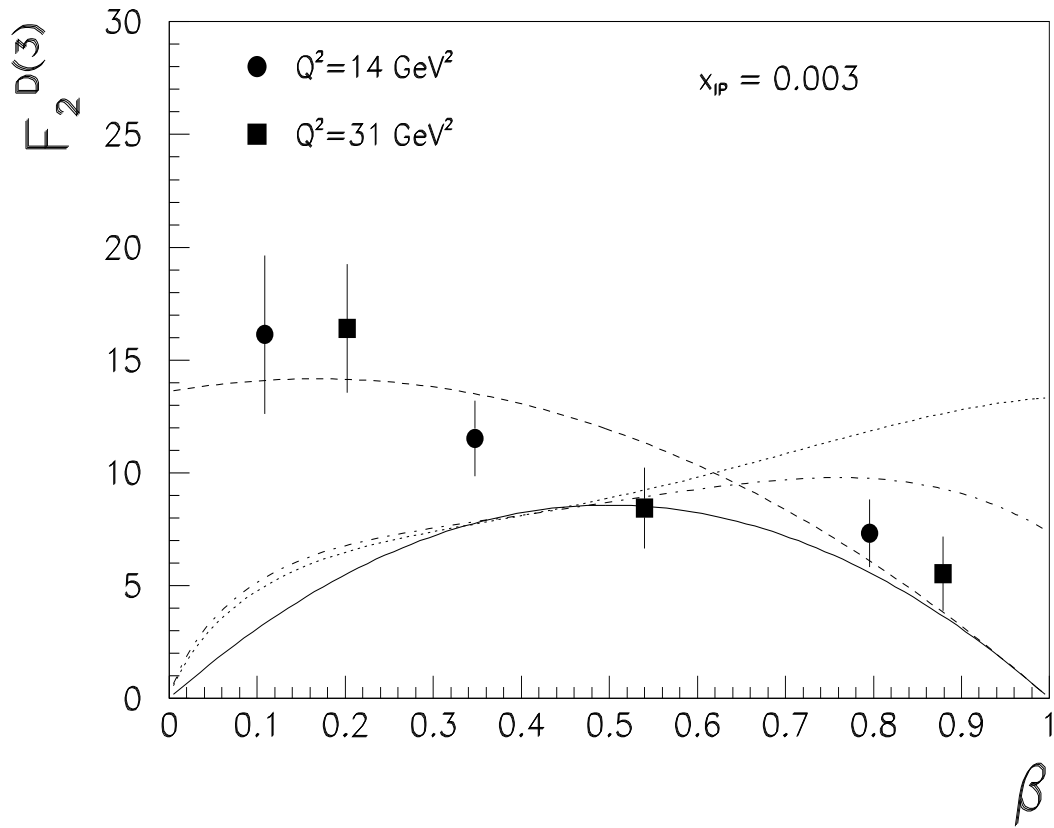}
\unitlength1cm
\begin{picture}(15,20)
\thicklines
\end{picture}
\caption{\label{f:fd3b}
        {    The diffractive structure function $F^{D(3)}_2$ as a function 
             of $\beta$ at $\xpom$ = 0.003. The error bars show the statistical
             and systematic errors added in quadrature. The full line
             shows the prediction of Hard-POMPYT. The dashed line shows the 
             prediction of Hard-POMPYT with an additional gluon contribution                suggested by the NZ model and
             fitted to the data (see text). The dashed-dotted (dotted) line 
             shows the prediction of the pomeron model based on the
             photon gluon fusion dynamics at 14 (31)~GeV$^2$ (see text).}}
\end{figure}

The dependence of $F^{D(3)}_2$ on $\beta$ in this analysis was determined as
follows:  the largest range in $\beta$ is covered for $\xpom=0.003$ as can
be seen in Fig.~\ref{f:fd3}. We chose, therefore, for every ($M_X$, $Q^2$)
bin that $F^{D(3)}_2$ value with $\xpom$ closest to 0.003 and determined by
interpolation with the expression $F^{D(3)}_2= (\xpom/0.003)^a\cdot
F^{D(3)}_2(\xpom)$ the value of $F^{D(3)}_2(Q^2,\beta,\xpom=0.003)$.  The
result is shown in Fig.~\ref{f:fd3b} as a function of $\beta$. Compared to
our previous measurement, the range in $\beta$ is considerably increased.
Figure~\ref{f:fd3b} shows that $F^{D(3)}_2$ rises as $\beta$ decreases. This
is expected from the QCD evolution of the parton densities in the pomeron. 

The $\beta$ dependence of $F^{D(3)}_2$ is sensitive to the dynamics of the
$\gamma^*$-pomeron interaction and can distinguish between different pomeron
models. In Fig.~\ref{f:fd3b} $F^{D(3)}_2$ is compared with the predictions
of various models. In the model of ~\cite{Ellisetal,Buch}, the pomeron is
represented by a single gluon leading to photon-gluon fusion followed by
subsequent colour compensation. The colour compensation is considered to be
sufficiently soft so that the dynamical properties of the photon-gluon
fusion process remain unchanged.  The predictions of ~\cite{Buch}, shown as
the dashed-dotted and dotted lines, were normalized to the value of
$F^{D(3)}_2$ at $\beta=0.5$, $\xpom=0.003$ and $Q^2=31$~GeV$^2$.  The model
fails to reproduce the rise of $F^{D(3)}_2$ towards small $\beta$ values and
the $Q^2$ dependence at large $\beta$. The prediction of the Hard-POMPYT
model (full line) where the $\gamma^*$-pomeron interaction results in a
quark-antiquark final state (called the hard component) and where the quark
momentum density of the pomeron is given by $\beta f(\beta) \propto
\beta\cdot (1-\beta)$ also fails to describe the measured $\beta$ dependence
of $F^{D(3)}_2$. However, agreement can be obtained by the inclusion of a
soft component in the pomeron leading to the form 
$$\beta f(\beta) \propto
\beta\cdot (1-\beta) + \frac{g}{2} \cdot (1-\beta)^2 $$ 
as suggested in the NZ model ~\cite{Nz}. A fit to the data yielded $g =0.78
\pm 0.32$ which is in agreement with the NZ prediction of $g \approx 1$. The
fit, shown as the dashed curve in Fig.~\ref{f:fd3b}, gives a good
description of the data. 

\section{Conclusions}

A novel method was used to extract the diffractive cross section in
deep-inelastic electron-proton scattering. Previous analyses were based on
pseudorapidity gap distributions and depended on the detailed modelling of
the diffractive and nondiffractive contributions.  The new method is based
on the measurement of the mass $M_X$ of the system $X$ resulting from the
dissociation of the virtual photon and assumes that, for nondiffractive
scattering, low $\ln M_X^2$ of the hadronic system observed in the detector
are exponentially suppressed. The exponential slope and thus the
nondiffractive contribution were obtained from the data and were found to be
independent of the specific form of the diffractive contribution. 

The $W$ dependence of the diffractive cross section
$d\sigma^{diff}_{\gamma^*p \to XN}(M_X,W,Q^2,M_N<4\, {\rm GeV})/dM_X$
measured at large $Q^2$ between $10$ and $56$~GeV$^2$ yielded a value of
$\overline{\alphapom} = 1.23 \pm 0.02 (stat) \pm 0.04 (syst)$ for the
pomeron trajectory averaged over $t$. The same $W$ dependence was found for
a subset of the events which have $M_X < 3$ GeV and which are characterized
by a large rapidity gap and a negligible nondiffractive background. The
value of $\overline{\alphapom}$ was obtained under the assumption that the
contribution from longitudinal photons is zero. Assuming $\sigma_L^{diff} =
(Q^2/M^2_X) \sigma_{T}^{diff}$ leads to a larger value of
$\overline{\alphapom} = 1.28 \pm 0.02 (stat)$.  The value for
$\overline{\alphapom}$ measured in this experiment is substantially larger
than the result found for the soft pomeron in hadron~- hadron scattering
averaged over $t$, $\overline{\alphapom} = 1.05$, a value which is also
consistent with data on $\rho^o$ production by $real$ photons at
HERA~\cite{Zgprho,H1prho}. The observation that $\overline{\alphapom}$ is
substantially larger in diffractive DIS than expected for the soft pomeron
suggests that in the kinematic region of this analysis a substantial part of
the diffractive DIS cross section originates from processes which can be
described by perturbative QCD.

\section*{Acknowledgements}          
The experiment was made possible by the inventiveness and the diligent   
efforts of the HERA machine group who continued to run HERA most 
efficiently during 1993.
                                                                               
The design, construction and installation of the ZEUS detector has             
been made possible by the ingenuity and dedicated effort of many people     
from inside DESY and from the home institutes, who are not listed as authors.
Their contributions are acknowledged with great appreciation.
                   
The strong support and encouragement of the DESY Directorate                   
has been invaluable. 

We also gratefully acknowledge the support of the DESY computing and network
services.

%

%

\begin{table}[ht]
\begin{center}
\begin{tabular}[t]{|c|c|r@{-}r|r|r|r@{$ $}r||r@{~$\pm$}r@{~$\pm$}r|} \hline

$M_X$&~$Q^2$&\multicolumn{2}{c|}{$W$}&{}&{}&
\multicolumn{2}{c||}{}&d$\sigma/dM_X$&stat&syst\\

range&range&\multicolumn{2}{c|}{range}&${\cal N}_{obs}$&${\cal N}^{e gas}$&
\multicolumn{2}{c||}{${\cal N}^{nondiff}$}&
\multicolumn{3}{c|}{}\\

(GeV)&(GeV$^2$)&\multicolumn{2}{c|}{(GeV)}&{}&{}&
\multicolumn{2}{c||}{}&
\multicolumn{3}{c|}{(nb/GeV)}\\ \hline \hline

$<$ 3  & 10-20 &  60&  74& 49& 0& 2& $\pm$ 2& 36.7 & 5.7 &$^{~5.9}_{10.3}$\\
       &       &  74&  90& 67& 0& 1& $\pm$ 1& 46.6 & 6.4 &$^{~6.1}_{~7.0}$\\
       &       &  90& 110& 87&24& 1& $\pm$ 1& 55.5 & 9.2 &$^{~8.3}_{12.8}$\\
       &       & 110& 134& 91& 8& 4& $\pm$ 2& 68.6 & 8.4 &$^{~7.5}_{~8.2}$\\
       &       & 134& 164& 97& 8& 1& $\pm$ 1& 78.4 & 9.0 &$^{11.0}_{11.8}$\\
       &       & 164& 200&118&24& 1& $\pm$ 1& 94.7 &11.3 &$^{13.3}_{16.1}$\\ \hline \hline
$<$ 3  & 20-56 &  60&  74& 19& 0& 5& $\pm$ 2&  4.6 & 1.4 &$^{~1.3}_{~1.3}$\\
       &       &  74&  90& 21& 0& 2& $\pm$ 1&  6.8 & 1.8 &$^{~1.4}_{~1.8}$\\
       &       &  90& 110& 28& 0& 3& $\pm$ 2&  8.1 & 2.0 &$^{~1.4}_{~1.4}$\\
       &       & 110& 134& 27& 0& 1& $\pm$ 1& 10.1 & 2.3 &$^{~1.1}_{~2.2}$\\
       &       & 134& 164& 28&16& 1& $\pm$ 1&  7.7 & 3.0 &$^{~4.2}_{~1.1}$\\
       &       & 164& 200& 42& 0&  &        & 16.3 & 3.2 &$^{~2.1}_{~3.1}$\\
       &       & 200& 245& 25& 0&  &        & 18.6 & 4.0 &$^{~2.2}_{~4.1}$\\ \hline \hline
 3-7.5  & 10-20 &  60&  74& 90& 0&37& $\pm$18& 35.4 & 8.6 &$^{11.3}_{18.4}$\\
        &       &  74&  90&110& 0&21& $\pm$ 9& 53.2 & 6.0 &$^{9.6}_{11.7}$\\
        &       &  90& 110&104& 0&13& $\pm$ 6& 71.0 & 6.9 &$^{~5.7}_{~9.2}$\\
        &       & 110& 134&108& 0&26& $\pm$ 8& 80.4 & 7.6 &$^{~5.6}_{~8.8}$\\
        &       & 134& 164&132& 0& 7& $\pm$ 3& 99.3 & 8.8 &$^{~6.0}_{~8.9}$\\
        &       & 164& 200&129& 0& 5& $\pm$ 3&106.5 & 9.8 &$^{~4.3}_{12.8}$\\
        &       & 200& 245& 78& 0& 1& $\pm$ 1&118.3 &11.5 &$^{~7.1}_{15.4}$\\ \hline \hline
 3-7.5  & 20-56 &  74&  90& 64& 0&24& $\pm$12& 12.8 & 3.2 &$^{~3.5}_{~5.1}$\\
        &       &  90& 110& 59& 0&26& $\pm$10& 13.7 & 2.8 &$^{~4.0}_{~2.9}$\\
        &       & 110& 134& 48& 0&13& $\pm$ 6& 15.9 & 2.5 &$^{~2.2}_{~1.8}$\\
        &       & 134& 164& 64&16& 5& $\pm$ 3& 20.4 & 4.1 &$^{~6.9}_{~1.8}$\\
        &       & 164& 200& 72& 0& 2& $\pm$ 1& 31.0 & 4.0 &$^{~2.5}_{~3.1}$\\
        &       & 200& 245& 54& 0&  &        & 32.0 & 4.4 &$^{~2.6}_{~2.9}$\\ \hline \hline
 7.5-15 & 10-20 & 134& 164&134& 0&47& $\pm$14& 57.5 & 6.9 &$^{~8.6}_{10.3}$\\
        &       & 164& 200& 85& 0&29& $\pm$ 9& 62.1 & 6.3 &$^{~8.1}_{11.8}$\\
        &       & 200& 245& 63& 0& 8& $\pm$ 4& 69.7 & 7.4 &$^{~7.0}_{13.2}$\\ \hline \hline
 7.5-15 & 20-56 & 164& 200& 77& 0&13& $\pm$ 6& 23.8 & 3.0 &$^{~1.7}_{~2.9}$\\
        &       & 200& 245& 52& 0& 3& $\pm$ 2& 26.9 & 3.4 &$^{~1.3}_{~3.2}$\\ \hline
\end{tabular}
\end{center}
\label{f2d}
\caption{Results for the diffractive scattering via $\gamma^* p \to X + N$,
where $N$ is the proton or dissociated nucleonic system with mass $M_N <$~4~GeV.
The table contains, for each bin, the ranges of $M_X$, $Q^2$, and $W$; 
the number of observed events in the bin, ${\cal N}_{obs}$; the number of background
events from electron gas scattering, ${\cal N}^{e gas}$ (obtained from the number of events found with unpaired electron bunches and scaled up by a factor of $\approx$ 8 to account for the difference in electron currents) and from nondiffractive
scattering, ${\cal N}^{nondiff}$; 
the value of the differential cross section $d\sigma/dM_X$ at the $Q^2$ values of 14 and 31 GeV$^2$ and the logarithmic means of the $W$ intervals averaged over the specified $M_X$ range and its statistical and
systematic errors.  The overall normalization uncertainty of 3.5\% is not included.
}
\end{table}

\begin{table}[ht]
\begin{center}
\begin{tabular}[t]{|c|r|r|r|r|} \hline
 
$Q^2$ (GeV$^2$)  $/\; M_X ({\rm GeV})$ &$ 2m_\pi-3$  &$  3-7.5$  &$ 7.5-15$  &$ 2m_\pi-15$  \\ \hline \hline
 
 14  & 2.9 $\pm$ 0.7 $\%$ &  4.8 $\pm$ 1.0 $\%$ &  4.6 $\pm$ 1.2 $\%$ & 12.3 $\pm$ 1.7 $\%$ \\ \hline
 31  & 0.9 $\pm$ 0.2 $\%$ &  2.7 $\pm$ 0.6 $\%$ &  3.4 $\pm$ 0.8 $\%$ & ~7.0 $\pm$ 1.0 $\%$ \\ \hline
\end{tabular}
\end{center}
\caption{The ratio $r^{diff} = \int dM_X \sigma^{diff}_{\gamma^*p \to XN}(M_N < 4~{\rm GeV}) / \sigma^{tot}_{\gamma^*p}$  integrated over different $M_X$ bins 
at $W$ = 181~GeV and  $Q^2$ = 14 GeV$^2$ and 31~GeV$^2$, 
respectively.
}
\end{table}

\clearpage

\clearpage

\clearpage
 
\clearpage

\clearpage

\clearpage

\clearpage

\clearpage

\clearpage

\clearpage

\clearpage

\clearpage

\clearpage

\clearpage

\end{document}